\documentclass[aps,prl,twocolumn,superscriptaddress]{revtex4-1}
\usepackage{graphicx}
\usepackage[hidelinks]{hyperref} 
\graphicspath{{./}}

\begin{document}


\title{Spiral spin-liquid and the emergence of a vortex-like state in MnSc$_2$S$_4$}

\author{Shang Gao}
\affiliation{Laboratory for Neutron Scattering and Imaging, Paul Scherrer Institut, CH-5232 Villigen PSI, Switzerland}
\affiliation{Department of Quantum Matter Physics, University of Geneva, CH-1211 Geneva, Switzerland}

\author{Oksana Zaharko}
\email[]{oksana.zaharko@psi.ch}
\affiliation{Laboratory for Neutron Scattering and Imaging, Paul Scherrer Institut, CH-5232 Villigen PSI, Switzerland}

\author{Vladimir Tsurkan}
\affiliation{Experimental Physics V, University of Augsburg, D-86135 Augsburg, Germany}
\affiliation{Institute of Applied Physics, Academy of Sciences of Moldova, MD-2028 Chisinau, Republic of Moldova}

\author{Yixi Su}
\affiliation{J\"ulich Center for Neutron Science JCNS-MLZ, Forshungszentrum J\"ulich GmbH, Outstation at MLZ, D-85747 Garching, Germany}

\author{Jonathan S. White}
\affiliation{Laboratory for Neutron Scattering and Imaging, Paul Scherrer Institut, CH-5232 Villigen PSI, Switzerland}

\author{Gregory S. Tucker}
\affiliation{Laboratory for Neutron Scattering and Imaging, Paul Scherrer Institut, CH-5232 Villigen PSI, Switzerland}
\affiliation{Laboratory for Quantum Magnetism, \'Ecole Polytechnique F\'ed\'erale de Lausanne, CH-1015 Lausanne, Switzerland}

\author{Bertrand Roessli}
\affiliation{Laboratory for Neutron Scattering and Imaging, Paul Scherrer Institut, CH-5232 Villigen PSI, Switzerland}

\author{Frederic Bourdarot}
\affiliation{CEA et Universit\'e Grenoble Alpes, INAC-MEM-MDN, F-38000 Grenoble, France}

\author{Romain Sibille}
\affiliation{Laboratory for Neutron Scattering and Imaging, Paul Scherrer Institut, CH-5232 Villigen PSI, Switzerland}
\affiliation{Laboratory for Scientific Developments and Novel Materials, Paul Scherrer Institut, CH-5232 Villigen PSI, Switzerland}

\author{Dmitry Chernyshov}
\affiliation{Swiss-Norwegian Beamlines at the European Synchrotron Radiation Facility, F-38000 Grenoble, France}

\author{Tom Fennell}
\affiliation{Laboratory for Neutron Scattering and Imaging, Paul Scherrer Institut, CH-5232 Villigen PSI, Switzerland}

\author{Alois Loidl}
\affiliation{Experimental Physics V, University of Augsburg, D-86135 Augsburg, Germany}

\author{Christian R\"uegg}
\affiliation{Laboratory for Neutron Scattering and Imaging, Paul Scherrer Institut, CH-5232 Villigen PSI, Switzerland}
\affiliation{Department of Quantum Matter Physics, University of Geneva, CH-1211 Geneva, Switzerland}

\date{\today}

\begin{abstract}
Spirals and helices are common motifs of long-range order in magnetic solids, and they may also be organized into more complex emergent structures such as magnetic skyrmions and vortices. A new type of spiral state, the spiral spin-liquid, in which spins fluctuate collectively as spirals, has recently been predicted to exist. Here, using neutron scattering techniques, we experimentally prove the existence of a spiral spin-liquid in MnSc$_2$S$_4$ by directly observing the `spiral surface' -- a continuous surface of spiral propagation vectors in reciprocal space. We elucidate the multi-step ordering behavior of the spiral spin-liquid, and discover a vortex-like triple-\textbf{q} phase on application of a magnetic field. Our results prove the effectiveness of the $J_1$--$J_2$ Hamiltonian on the diamond lattice as a model for the spiral spin-liquid state in MnSc$_2$S$_4$, and also demonstrate a new way to realize a magnetic vortex lattice.

\end{abstract}

\pacs{}

\maketitle


Magnetic frustration, where magnetic moments (spins) are coupled through competing interactions that cannot be simultaneously satisfied \cite{balents_2010}, usually leads to highly cooperative spin fluctuations \cite{bramwell_2001,Henley_2010} and unconventional long-range magnetic order \cite{reimers_1991,tchernyshyov_2002}. An archetypal ordering in the presence of frustration is the spin spiral.  Competing interactions and spiral orders give rise to many phenomena in magnetism, including the multitudinous magnetic phases of rare earth metals~\cite{Jensen:1991ux}, domains with multiferroic properties \cite{cheong_2007,mostovoy_2006}, and topologically non-trivial structures such as the emergent skyrmion lattice \cite{nagaosa_2013,muhlbauer_2009}. 

\begin{figure*}
\includegraphics[width=0.90\textwidth]{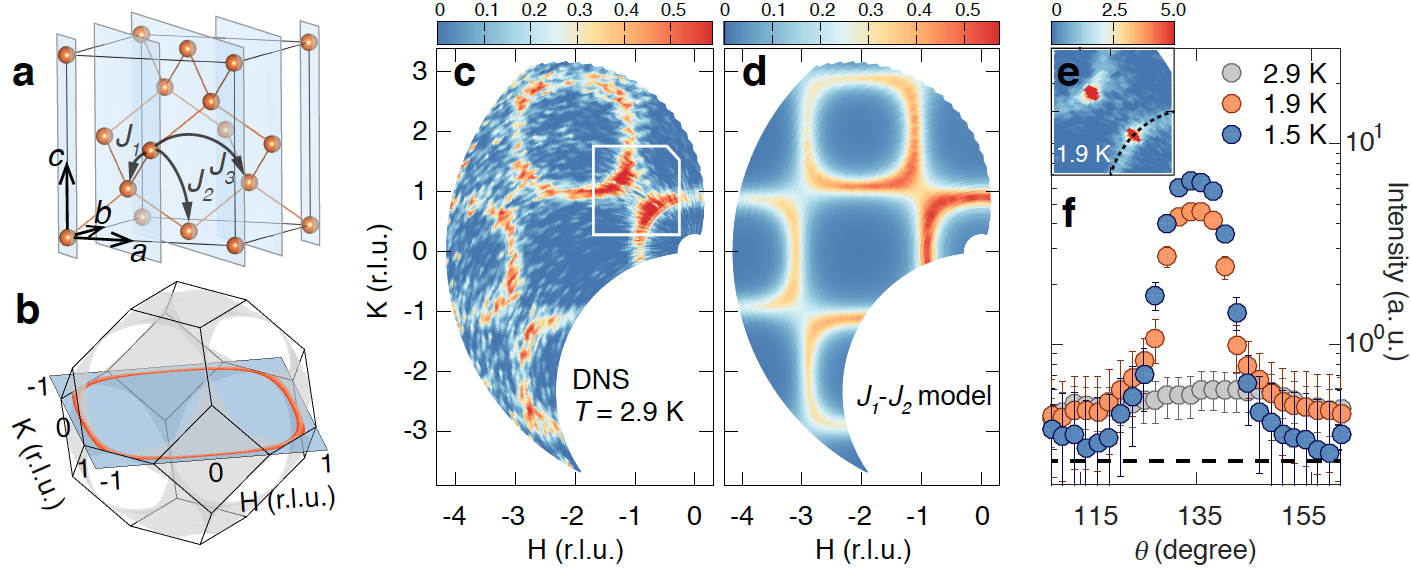}
\caption{\textbf{Spiral spin-liquid. (a)} Diamond lattice of Mn$^{2+}$ ions in MnSc$_2$S$_4$, (110) planes are shaded blue. \textbf{(b)} Spiral surface (gray) predicted by mean-field theory for the $J_1$-$J_2$ model with the ratio $|J_2/J_1|=0.85$. The red ring emphasizes a cut within the (HK0) plane (blue). \textbf{(c)}. Diffuse scattering intensities in the (HK0) plane measured at 2.9 K. The white square outlines the area shown in panel e. \textbf{(d)} Monte Carlo simulations for spin correlations in the (HK0) plane using the $J_1$-$J_2$ model with the ratio $|J_2/J_1|=0.85$ and $T/|J_1|=0.55$. \textbf{(e)} Diffuse scattering intensities around (110) (contour outlined in panel c) measured at 1.9 K, showing the coexistence of Bragg peaks and the diffuse signal. The dashed arc describes the path for the 1D cut presented in panel f, where the polar angle $\theta$ is used as the $x$ axis. \textbf{(f)} Comparision of the 1D cut at $T=2.9, 1.9,$ and 1.5 K, showing the diffuse signal extends down to the base temperature of 1.5 K. The horizontal dashed line indicates the background level averaged from 1D cuts with varying radius. \label{fig:DNS}}
\end{figure*}

Recently, a new spiral state -- a spiral spin-liquid in which the ground states are a massively degenerate set of coplanar spin spirals -- was predicted to exist in the $J_1$--$J_2$ model on the diamond lattice (see Fig.~\ref{fig:DNS}a) \cite{bergman_2007,lee_2008,savary_2011}.  Although the diamond lattice is bipartite, and therefore unfrustrated at the near-neighbour (\textit{J$_{1}$}) level, the second-neighbour coupling (\textit{J$_{2}$}) can generate strong competition. For classical spins, mean-field calculations show that when $|J_{2}/J_{1}|>0.125$ the spiral spin-liquid appears, and that it is signified by an unusual continuous surface of propagation vectors \textbf{q} in reciprocal space (see Fig.~\ref{fig:DNS}b for the spiral surface of $|J_{2}/J_{1}|=0.85$).  At finite temperature, thermal fluctuations might select some specific \textbf{q}-vectors on the spiral surface \cite{bergman_2007}, resulting in an order-by-disorder transition \cite{villain_1980, henley_1989}. 

Until now, several series of $A$-site spinels, in which the magnetic $A$ ions form a diamond lattice, have been investigated, including: the cobaltates Co$_{3}$O$_{4}$ and CoRh$_{2}$O$_{4}$ \cite{suzuki_2007}; the aluminates \textit{M}Al$_{2}$O$_{4}$ with \textit{M} = Fe, Co, Mn \cite{tristan_2005,krimmel_2009,zaharko_2011, macdougall_2011}; and the scandium thiospinels \textit{M}Sc$_{2}$S$_{4}$ with \textit{M} = Fe, Mn \cite{fritsch_2004}. For the spinels with Fe$^{2+}$ at the $A$-site, the $e_{g}$ orbital angular momentum of Fe$^{2+}$ is active, making the pure spin $J_{1}$--$J_{2}$ model inadequate \cite{chen_2009}. Among the other compounds, CoAl$_{2}$O$_{4}$ and MnSc$_{2}$S$_{4}$ manifest the strongest frustration. For CoAl$_{2}$O$_{4}$, the ratio of $|J_{2}/J_{1}|$ has been identified as 0.109 \cite{zaharko_2011}, which is near, but still lower than, the 0.125 threshold for the spiral spin-liquid state.  Many experimental studies of MnSc$_{2}$S$_{4}$ \cite{fritsch_2004, krimmel_2006, mucksch_2007, giri_2005, buttgen_2006,kalvius_2006} suggest its relevance to the spiral spin-liquid, but the spiral surface has not been observed.

In this article, we present the results of extensive experimental studies of high-quality single crystals of MnSc$_2$S$_4$ by neutron scattering.  Our diffuse scattering data uncovers the spiral surface, providing direct evidence for the proposed spiral spin-liquid state.  The spiral spin-liquid enters a helical long-range ordered state through a multi-step ordering process, which we suggest is due to dipolar and third-neighbour ($J_3$) perturbations, which we quantify. Finally, by applying a magnetic field, an emergent triple-\textbf{q} phase is discovered. The \textbf{q} combination rule of this field-induced phase is similar to that observed in the skyrmion lattice \cite{nagaosa_2013, muhlbauer_2009}, which suggests $A$-site spinels are new candidate systems to realize vortex lattice states \cite{okubo_2012,kamiya_2014,wang_2015,rousochatzakis_2016}.

\begin{figure}
\includegraphics[width=0.45\textwidth]{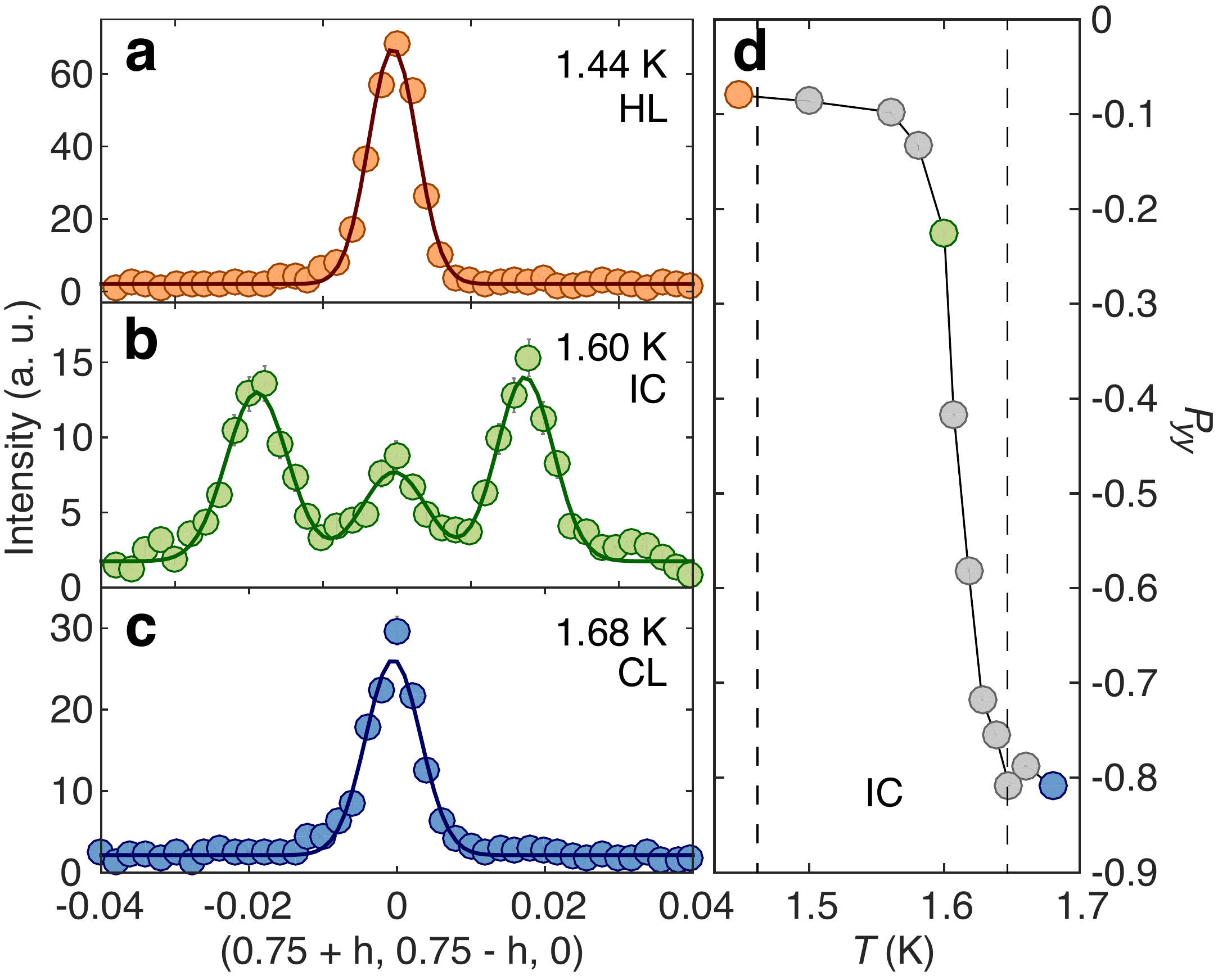}
\caption{\textbf{Multi-step ordering towards the helical ground state.} \textbf{(a-c)} Neutron diffraction data along the [1\=10] direction for the (0.75 0.75 0) reflection measured at 1.44, 1.60, and 1.68 K during cooling. HL, IC, and CL represent the helical, incommensurate, and sinusoidally-modulated collinear phases, respectively. \textbf{(d)} Temperature dependence of the SNP element $P_{yy}$ measured with the (HHL) plane horizontal. Orange, green, and blue points correspond to $T=1.44, 1.60,$ and 1.68 K, respectively. \label{fig:CryoPAD}}
\end{figure}

\begin{figure*}
\includegraphics[width=0.9\textwidth]{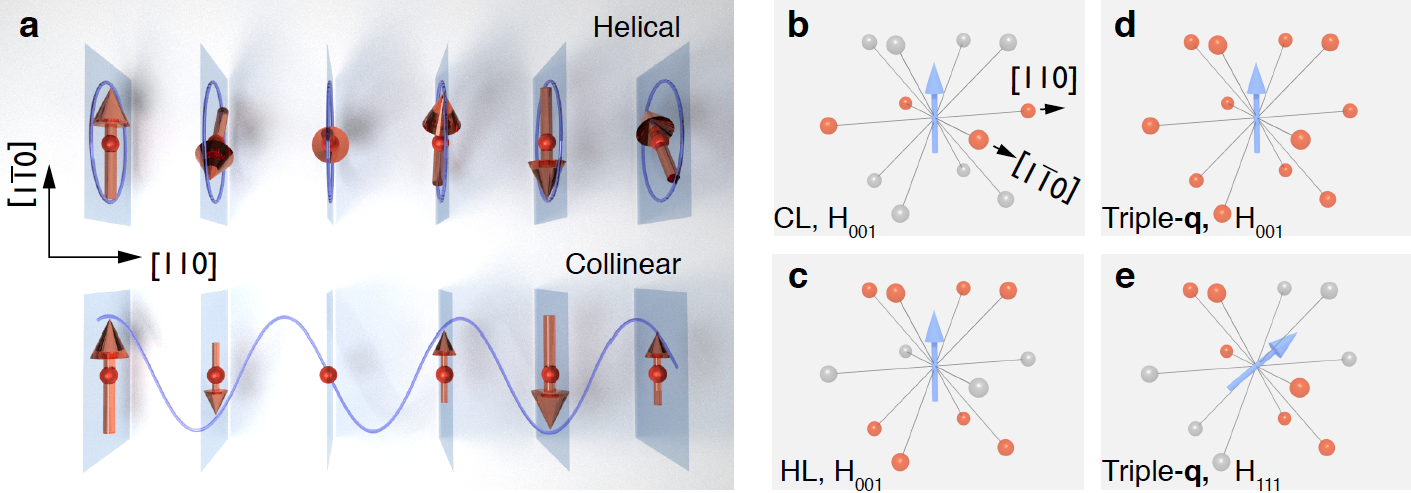}
\caption{\textbf{Spin structures and their magnetic field response}. \textbf{(a)} The helical structure (top) and the sinusoidally-modulated collinear structure (bottom) refined from 1.38 and 1.70 K datasets with $R_{f2}$ = 0.14 and 0.17, respectively. Blue shaded planes are perpendicular to the propagation vector (\textit{e.g.}, (110) planes for \textbf{q} = (0.75 0.75 0), which are also shown in the crystal structure of Fig.~\ref{fig:DNS}a); over these planes the ordered moments have the same size and orientation. For the helical structure, the moment is refined to be 5.27(23) $\mu_B$. For the collinear structure, the moment is sinusoidally modulated from 0 to 4.77(20) $\mu_B$. \textbf{(b-e)} Intensity distribution for the 12 arms of the $\langle$0.75 0.75 0$\rangle$ star under magnetic field. They are extracted from refinements of neutron diffraction datasets. Orange spheres indicate arms with nearly equal intensites; grey spheres indicate arms with zero intensities; and blue arrows indicate the field directions. Panel b shows the distribution in the collinear phase measured at $T = 1.80$ K under a [001] magnetic field of 3.5 T. Panel c shows the expected distribution in the helical phase under a [001] magnetic field, and the disappearance of the arms in the (001) plane are confirmed experimentally for $H$ below $\sim3$ T. Panel d shows the distribution in the triple-\textbf{q} phase measured at $T = 1.30$ K under a  3.5 T magnetic field along [001]. Panel e shows the distribution in the triple-\textbf{q} phase measured at $T = 1.60$ K under a 3.5 T magnetic field along [111].
\label{fig:structures}}
\end{figure*}

\begin{figure}
\includegraphics[width=0.48\textwidth]{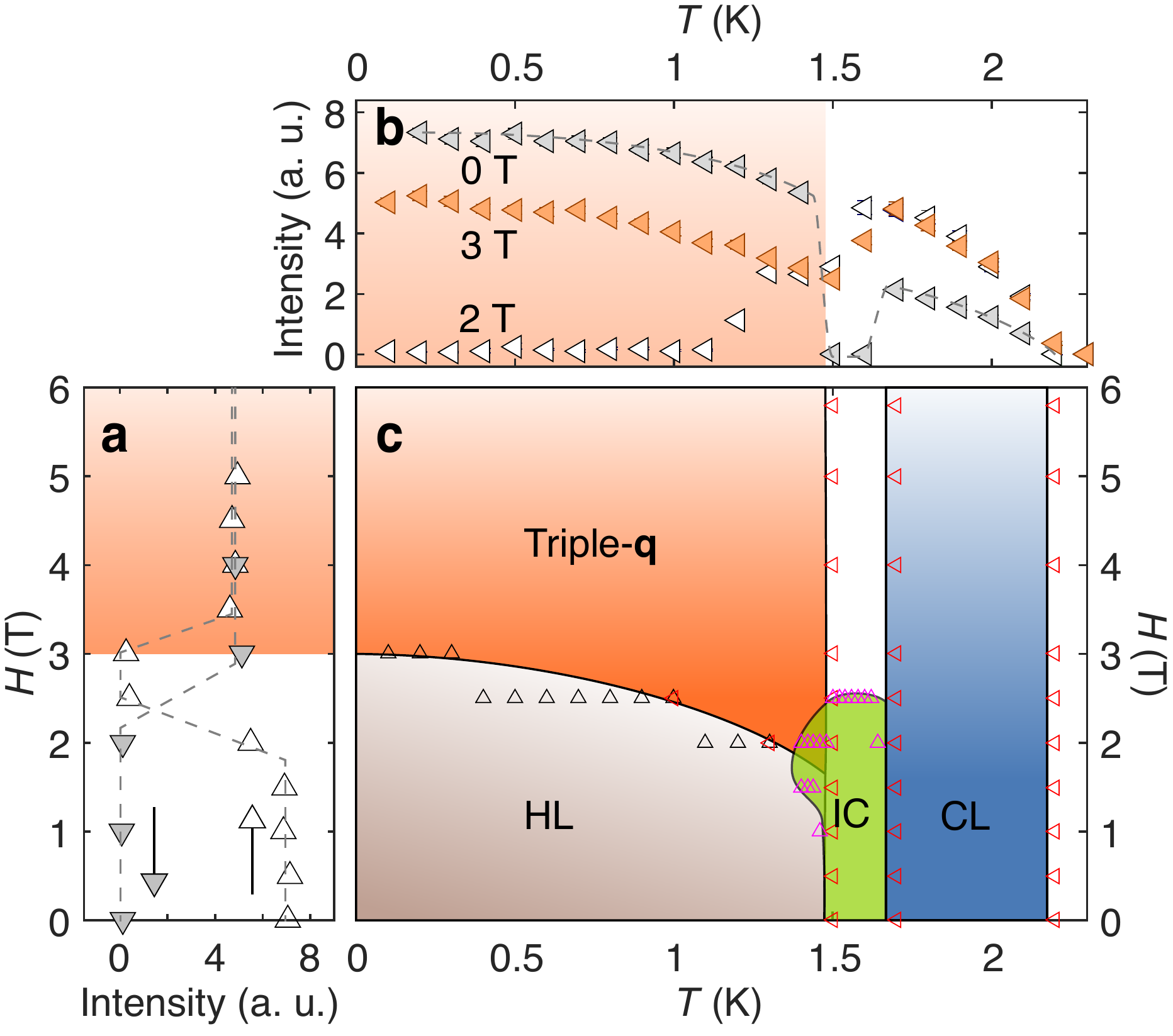}
\caption{\textbf{Phase diagram of MnSc$_2$S$_4$ under a magnetic field along the [001] direction.} \textbf{(a)} Field dependence of the intensity of the (0.75 0.75 0) reflection measured at $T=0.1$ K. Up-pointing empty (down-pointing filled) triangles belong to the measurements with increasing (decreasing) field. \textbf{(b)} Temperature dependence of the neutron diffraction intensity of the (0.75 0.75 0) reflection measured under different magnetic fields. \textbf{(c)} Phase diagram from neutron diffraction experiments. Up-pointing (left-pointing) triangles mark the transition positions extracted from the measurements with increasing field (decreasing temperature). HL, CL and IC represent the helical, sinusoidally-modulated collinear, and incommensurate phases, respectively. The IC phase disappears under field cooling.
\label{fig:phase_diagram}}
\end{figure} 

Neutron diffuse scattering measures the reciprocal space distribution of the quasi-static spin correlations and provides direct information on the ground state degeneracy in frustrated systems \cite{reimers_1992, Fennell_2009}. Fig.~\ref{fig:DNS}c presents our diffuse scattering results for the (HK0) plane of MnSc$_2$S$_4$, measured at 2.9 K. Strong intensities are observed near the Brillouin zone boundaries, forming a squared-ring pattern. This pattern, with its large span and non-ellipsoidal form in reciprocal space, is very different from those observed in less-frustrated systems such as CoAl$_{2}$O$_{4}$ \cite{macdougall_2011,zaharko_2011}, where intensities center closely around magnetic Bragg points that characterize incipient order. According to mean-field calculations \cite{bergman_2007}, a larger ratio of $|J_{2}/J_{1}|$ causes increased frustration and is accompanied by the expansion of the spiral surface in reciprocal space, expanding the ring in the (HK0) plane towards the Brillouin zone boundary. Thus the observed squared-ring feature affirms that MnSc$_{2}$S$_{4}$ is strongly frustrated. 

Monte Carlo simulations of the $J_1$--$J_2$ model on a $10\times10\times10$ superlattice can be directly compared with the experiment. The simulated neutron scattering pattern for the ratio $|J_2$/$J_1|=0.85$ and $T/|J_1|=0.55$ is shown in Fig.~\ref{fig:DNS}d. In Ref.~\cite{bergman_2007} this ratio of $|J_2/J_1|$ was shown to be a highly frustrated case of the Hamiltonian, with the spiral spin-liquid as its ground state. Thus the experimentally observed diffuse scattering directly proves the existence of the spiral surface and the spiral spin-liquid state in MnSc$_2$S$_4$. 

Previous studies revealed a transition into a commensurate long-range ordered state with $\mathbf{q} = (0.75\ 0.75\ 0)$ at $T\sim 2.3$ K \cite{fritsch_2004, krimmel_2006}. Fig.~\ref{fig:DNS}e shows the data at 1.9 K for the region around (--1 1 0).  Two strong magnetic Bragg peaks have appeared at (--0.75 0.75 0) and (--1.25 1.25 0), but a trace of the spiral surface is still discernible below the long-range ordering transition.  1D cuts along the arc of the spiral surface at different temperatures are compared in Fig.~\ref{fig:DNS}f, revealing the coexistence of long-range and short-range correlations down to 1.5 K (the base temperature of the diffuse scattering experiment). Such a coexistence signals the presence of strong fluctuations in the ordered phase and calls for a detailed study of the ordering behavior in MnSc$_2$S$_4$.

Using single crystal neutron diffraction, which is a direct probe of the long-range ordered state, we investigated the ordering to lower temperatures.  Fig.~\ref{fig:CryoPAD}a-c present scans along the [1\=10] direction for the (0.75 0.75 0) peak at $T=1.68, 1.60, 1.44$ K. In the region $1.64 >T>1.46$ K, an incommensurate (IC) phase with the propagation vector $\textbf{q}'=(0.75 \pm 0.02\ 0.75 \mp 0.02\ 0)$ was discovered, suggesting that the ordering of the spiral spin-liquid involves a multi-step process.

This multi-step character is also evident in our spherical neutron polarimetry SNP experiments \cite{brown_2006}. Fig.~\ref{fig:CryoPAD}d plots the temperature dependence of the $P_{yy}$ polarization element of the (0.75 0.75 0) peak, which describes the difference of the magnetic structure factors $M$ along $\hat{y}$/$\hat{z}$ directions ($\hat{x}$ along [\=1\=10], $\hat{y}$ along [001], and $\hat{z}$ along [\=110] for (HHL) plane horizontal): $P_{yy}= (M_yM_y^*-M_zM_z^*)/(M_yM_y^*+M_zM_z^*)$. The evolution of $P_{yy}$ from $-0.8$ to $-0.1$ with decreasing $T$ unambiguously reveals that the two commensurate phases for $T>1.64$ K and $T<1.46$ K possess different magnetic structures, even though their \textbf{q} vectors are exactly the same.

To determine the magnetic structures, complete sets of magnetic Bragg peak intensities were collected by neutron diffraction at $T=$ 1.38 and 1.70 K and  compared with theoretical models. The previously proposed cycloidal structure \citep{krimmel_2006} failed to fit either dataset. Instead, at 1.38 K a helical structure (Fig.~\ref{fig:structures}a, top), and at 1.70 K a sinusoidally-modulated collinear structure (Fig.~\ref{fig:structures}a, bottom) were found to successfully reproduce the datasets. The transition from the collinear structure to the helical structure occurs through the incommensurate phase, whose narrow temperature window is consistent with the continuous growth of the SNP $P_{yy}$ element for $1.64>T>1.46$ K.

According to previous theoretical studies \citep{bergman_2007,lee_2008}, the inclusion of an antiferromagnetic $J_3$ perturbation in the $J_1$--$J_2$ model selects the \textbf{q} position for the spiral but leaves the spiral plane undetermined; whether the ground state is cycloidal or helical depends on further anisotropic perturbations. Here, the observed helical structure provides strong evidence for the effect of dipolar interactions, which have been shown to favor the helical structure over the cycloidal one  \citep{lee_2008}.  Previous calculations reveal that a value of $J_3\sim-0.04$ K brings \textbf{q} to (0.75 0.75 0)\citep{bergman_2007,lee_2008}, and the dipolar interaction is $\sim$ 0.026 K for nearest neighbours.  However, our mean-field calculations reveal that long-range order at \textbf{q} = (0.75 0.75 0) occurs even in the case where $J_3$ equals zero and only the dipolar perturbation exists.  Dipolar interactions might also contribute to the appearance of the collinear phase. By decomposing the $\textbf{q}$ = (0.75 0.75 0) helicoid into two sinusoidally-modulated collinear structures with spins along the [1\=10] and [001] directions, we find that the dipolar interactions favor the [1\=10] one by an energy difference of 0.003 K per spin. This in-plane anisotropy stabilizes the helicoid in-plane component and helps the development of the observed collinear phase \cite{mostovoy_2006, mochizuki_2009}. We note that all these perturbations are small compared to $J_1$ and $J_2$, such that the spiral spin-liquid we observed appear to be quite ideal, as evidenced by the close agreement between $J_1$--$J_2$ theory and experiment in Fig.~\ref{fig:DNS}c,d.

The established collinear and helical phases are both single-\textbf{q} structures, meaning that the 12 arms of the $\langle$0.75 0.75 0$\rangle$ star form independent magnetic domains. Fig.~\ref{fig:structures}b,c summarize the response of all the 12 arms for the collinear and helical phases when we applied a magnetic field along the [001] direction. As is discussed in the supplementary information, the observed intensity distributions are consistent with the anisotropic susceptibility effect expected for magnetic domains and confirm the single-\textbf{q} character of the collinear and helical phases.

However, after cooling from the collinear phase in a field of 3.5 T, refinements of the neutron diffraction datasets reveal that although the single-arm structure stays the same, the previously suppressed arms reappear with all the 12 arms having about the same intensities. This intensity re-distribution, which is summarized in Fig. \ref{fig:structures}d, is contradictory to the domain effect expected for a single-\textbf{q} structure, and evidences the emergence of a field-induced multi-\textbf{q} phase. By studying the $H$ and $T$ dependence of the intensity of the (0.75 0.75 0) Bragg peak (typical scans are shown in Fig.~\ref{fig:phase_diagram}a,b), the extent of this multi-\textbf{q} phase is mapped out in Fig.~\ref{fig:phase_diagram}c.

We propose the field-induced phase to be a triple-\textbf{q} state with:
\begin{equation}
\mathbf{M}(\mathbf{r})=\sum_{j=1}^3( \mathbf{m}_je^{i( \mathbf{q}_j \cdot \mathbf{r}+\phi_j)}+ \mathrm{c.\thinspace c.}),
\end{equation}
where the three coplanar $\textbf{q}_j$ satisfy $\sum_{j=1}^3 \mathbf{q}_j=0$ (\textit{e.g.}, $\textbf{q}_1$ = (0.75 0.75 0), $\textbf{q}_2$ = (0 $\overline{0.75}$ 0.75), and $\textbf{q}_3$ = $(\overline{0.75}$ 0 $\overline{0.75})$), similar to that observed in the skyrmion lattice \cite{muhlbauer_2009}; $\mathbf{m}_j$ is the real basis vector perpendicular to \textbf{q}$_j$; $\phi_j$ describes an additional phase factor; and c.c. is the complex conjugate. Four triple-\textbf{q} domains formed in this way are symmetrically equivalent under the [001] magnetic field, which explains the equal intensity distribution shown in Fig.~\ref{fig:structures}d. In contrast, in Fig.~\ref{fig:structures}e we plot the distribution measured under a 3.5 T magnetic field along the [111] direction. This [111] field breaks the symmetry and therefore only one triple-\textbf{q} domain with arms perpendicular to $H$ can be observed.

Neutron diffraction is not sensitive to the phase factor $\phi_j$. In the supplementary information, we present two possible structures that preserve the $C_3$ symmetry along the [111] direction together with the expected structures with a shorter \textbf{q} = (0.25 0.25 0) that might be realized in less frustrated $A$-site spinels \cite{bergman_2007}. In these structures, spin components in the (111) plane exhibit a winding behavior around the $C_3$ axis, resulting in a vortex state similar to that predicted in frustrated antiferromagnets \cite{okubo_2012,kamiya_2014}. The relatively large wavevector/short pitch of the spirals makes these structures much smaller than currently known skyrmions \cite{nagaosa_2013}. Considering that the winding feature persists regardless of the choice of $\phi_j$, our experiments reveal the $A$-site spinels to be a new system to realize an atomic-sized vortex lattice.

In theory, thermal fluctuations select an ordered state with $\mathbf{q}\sim(0\ 0\ 1)$ for the spiral spin-liquid by an order-by-disorder mechanism \cite{bergman_2007}. The multi-step ordering process and wavevector we have observed show clearly that MnSc$_2$S$_4$ does not support an order-by-disorder transition. Our studies show that for the prospect of realizing an order-by-disorder transition, it is crucial to identify materials with reduced perturbations from the dipolar and further-neighbour exchange interactions. However, it is clear that fluctuations play an essential, but not understood, role in MnSc$_2$S$_4$.  The collinear phase retains considerable spin fluctuations, as manifested in the amplitude modulation of its ordered moments.  We speculate that the progressive suppression of these fluctuations on cooling could set up a temperature dependent dipolar interaction whose changing competition with the exchange interactions could be at the origin of the incommensurate phase.  On the other hand, the search for spiral spin-liquids and order-by-disorder transition should not be restricted to the $A$-site spinels or even to the diamond lattice. The primary constituent to realize spiral spin-liquids is the frustrated $J_1$--$J_2$ model, which can be constructed for most bipartite lattices. One recent example is the honeycomb lattice, where a two-dimensional analogue of the spiral spin-liquid state and a subsequent order-by-disorder transition have been predicted \cite{mulder_2010}.

In summary, by neutron diffraction and diffuse scattering, we confirm the existence of a spiral spin-liquid in MnSc$_2$S$_4$, which exhibits the most frustrated ratio of $|J_2/J_1|=0.85$ in the known $A$-site spinels. At lower temperature, a multi-step ordering  process of the spiral spin-liquid is observed,  reflecting the significance of dipolar interactions in MnSc$_2$S$_4$. Under magnetic field, an emergent vortex-like triple-\textbf{q} phase is discovered, which establishes the $A$-site spinel as a promising system to realize vortex lattice states.

\section{Methods}
\textbf{Crystal growth and synchrotron X-ray experiments.} MnSc$_{2}$S$_{4}$ single crystals were grown with the chemical transport technique. Single crystal synchrotron X-ray diffraction experiment were performed on the Swiss-Norwegian Beamline SNBL BM01 at the European Synchrotron Radiation Facility (Grenoble, France). Details for the refinements are presented in the supplements. 

\textbf{Neutron diffuse scattering experiments.} Neutron diffuse scattering experiments were performed on the high-flux time-of-flight spectrometer DNS at MLZ (Garching, Germany) with a coaligned sample of $\approx 30$ mg. The pyrolytic graphite PG(002) monochromator together with Be-filter were used to select the incoming neutron wavelength of 4.5 \AA. To obtain the magnetic diffuse scattering signal, the intensities in the X spin-flip (SF) channel were measured, where the X direction is in the horizontal plane. The X-SF measurements at 50 K were subtracted as the background.

\textbf{Spherical neutron polarimetry experiments.} The polarized neutron diffraction experiments were performed on the cold neutron triple-axis spectrometer TASP with MuPAD at the spallation neutron source SINQ of Paul Scherrer Institut (Villigen, Switzerland) and the thermal neutron triple-axis spectrometer IN22 with CRYOPAD at the Institut Laue-Langevin (Grenoble, France). For the TASP measurements, PG(002) monochromator and PG(002) analyzer selected the neutron wavelength of 3.19 \AA. The obtained  matrices are presented in the supplementary information. For the IN22 measurements, the standard Heusler-Heusler configuration with a fixed neutron wavelength of 2.36 \AA\ was used. A PG-filter was inserted before the sample.

\textbf{Single crystal neutron diffraction experiments.} Neutron diffraction datasets were collected on the thermal-neutron diffractometer TriCS at SINQ of Paul Scherrer Insitut (Villigen, Switzerland). Incoming neutron wavelengths of 1.18 \AA\ (Ge(311) monochromator) and 2.32 \AA\ (PG monochromator) were used for the measurements. Under zero field, 135 and 62 reflections were collected at $T = 1.70$ and 1.35 K, respectively. 131, 50, and 56 reflections were collected at $H_{001} = 3.5$ T, $T = 1.38$, 1.62, and 1.85 K, respectively. Finally, 67 reflections were collected at $H_{111} = 3.5$ T, $T = 1.60$ K. Refinements were carried out using FULLPROF \cite{rodriguez_1993}.

\textbf{Mapping of the \emph{H}-\emph{T} phase diagram.} The mapping of the \textit{H}-\textit{T} phase diagram was performed on TASP. A cryomagnet together with a dilution refrigerator was used. PG(002) monochromator and PG(002) analyzer selected the neutron wavelength of 4.22 \AA. A PG-filter was installed in between the sample and the analyzer. The crystal was mounted with the (HK0) plane horizontal. The intensity of the (0.75 0.75 0) reflection was carefully studied in cooling/warming processes under constant field and also in field-sweeping processes at constant temperatures. Reversibility across the phase boundary was used to separate the intrinsic phase transitions from domain effects.

\textbf{Monte Carlo simulations.} Monte Carlo simulations were performed using the ALPS package \cite{bauer_2011}. For the comparison with the DNS results, a $10\times10\times10$ superlattice was used. Calculations for the magnetic structure factor were averaged over 25000 sweeps after 10000 sweeps of thermalizations. 

\textbf{Mean-field calculations.} Interaction matrices were Fourier transformed to the reciprocal space and then diagonalized. Positions with the maximal eigenvalue produce the long-range-order \textbf{q} vectors at the mean-field approximation level. The dipolar interaction, when included, was cut beyond the distance of 5 unit cells.

\section{Acknowledgements}
We acknowledge helpful discussions with L. Balents, B. Normand,  J. H. Chen, A. Scaramucci, A. Cervellino, S. T{\'o}th, S. Ward, and M. Ruminy.
Our neutron scattering experiments were performed at the Swiss Spallation Neutron Source SINQ, Paul Scherrer Institut, Villigen, Switzerland, the Heinz Maier-Leibnitz Zentrum MLZ, Garching, Germany, and the Institut Laue-Langevin ILL, Grenoble, France. The magnetization measurements were carried out in the Laboratory for Scientific Developments and Novel Materials, Paul Scherrer Institut, Villigen, Switzerland.
This work was supported by the Swiss National Science Foundation under Grants No. 20021-140862, No. 20020-152734, and the SCOPES project No. IZ73Z0-152734/1. Our work was additionally supported by the Swiss State Secretariat for Education, Research and Innovation (SERI) through a CRG-grant and via the Deutsche Forschungsgemeinschaft by the Transregional Collaborative Research Center TRR 80. 
\section{Author contributions}
O. Z. and Ch. R. designed and supervised the project. V. T. prepared the single crystals. Sh. G. and O. Z. performed the experiments with Y. X. S. as the local contact for DNS, J. S. W. and G. S. T. for TASP, B. R. for MuPAD, F. B. for CRYOPAD, D. C. for SNBL, and R. S. for the SQUID measurements. Sh. G. and O. Z. performed the calculations and analyzed the data. The manuscript was written by Sh. G., O. Z., T. F., and Ch. R with input from all co-authors.

\bibliography{MSS_arXiv}

\begin{thebibliography}{38}%
\makeatletter
\providecommand \@ifxundefined [1]{%
 \@ifx{#1\undefined}
}%
\providecommand \@ifnum [1]{%
 \ifnum #1\expandafter \@firstoftwo
 \else \expandafter \@secondoftwo
 \fi
}%
\providecommand \@ifx [1]{%
 \ifx #1\expandafter \@firstoftwo
 \else \expandafter \@secondoftwo
 \fi
}%
\providecommand \natexlab [1]{#1}%
\providecommand \enquote  [1]{``#1''}%
\providecommand \bibnamefont  [1]{#1}%
\providecommand \bibfnamefont [1]{#1}%
\providecommand \citenamefont [1]{#1}%
\providecommand \href@noop [0]{\@secondoftwo}%
\providecommand \href [0]{\begingroup \@sanitize@url \@href}%
\providecommand \@href[1]{\@@startlink{#1}\@@href}%
\providecommand \@@href[1]{\endgroup#1\@@endlink}%
\providecommand \@sanitize@url [0]{\catcode `\\12\catcode `\$12\catcode
  `\&12\catcode `\#12\catcode `\^12\catcode `\_12\catcode `\%12\relax}%
\providecommand \@@startlink[1]{}%
\providecommand \@@endlink[0]{}%
\providecommand \url  [0]{\begingroup\@sanitize@url \@url }%
\providecommand \@url [1]{\endgroup\@href {#1}{\urlprefix }}%
\providecommand \urlprefix  [0]{URL }%
\providecommand \Eprint [0]{\href }%
\providecommand \doibase [0]{http://dx.doi.org/}%
\providecommand \selectlanguage [0]{\@gobble}%
\providecommand \bibinfo  [0]{\@secondoftwo}%
\providecommand \bibfield  [0]{\@secondoftwo}%
\providecommand \translation [1]{[#1]}%
\providecommand \BibitemOpen [0]{}%
\providecommand \bibitemStop [0]{}%
\providecommand \bibitemNoStop [0]{.\EOS\space}%
\providecommand \EOS [0]{\spacefactor3000\relax}%
\providecommand \BibitemShut  [1]{\csname bibitem#1\endcsname}%
\let\auto@bib@innerbib\@empty
\bibitem [{\citenamefont {Balents}(2010)}]{balents_2010}%
  \BibitemOpen
  \bibfield  {author} {\bibinfo {author} {\bibfnamefont {L.}~\bibnamefont
  {Balents}},\ }\href {http://dx.doi.org/10.1038/nature08917} {\bibfield
  {journal} {\bibinfo  {journal} {Nature}\ }\textbf {\bibinfo {volume} {464}},\
  \bibinfo {pages} {199} (\bibinfo {year} {2010})}\BibitemShut {NoStop}%
\bibitem [{\citenamefont {Bramwell}\ and\ \citenamefont
  {Gingras}(2001)}]{bramwell_2001}%
  \BibitemOpen
  \bibfield  {author} {\bibinfo {author} {\bibfnamefont {S.~T.}\ \bibnamefont
  {Bramwell}}\ and\ \bibinfo {author} {\bibfnamefont {M.~J.~P.}\ \bibnamefont
  {Gingras}},\ }\href {\doibase 10.1126/science.1064761} {\bibfield  {journal}
  {\bibinfo  {journal} {Science}\ }\textbf {\bibinfo {volume} {294}},\ \bibinfo
  {pages} {1495} (\bibinfo {year} {2001})}\BibitemShut {NoStop}%
\bibitem [{\citenamefont {Henley}(2010)}]{Henley_2010}%
  \BibitemOpen
  \bibfield  {author} {\bibinfo {author} {\bibfnamefont {C.~L.}\ \bibnamefont
  {Henley}},\ }\href {\doibase doi:10.1146/annurev-conmatphys-070909-104138}
  {\bibfield  {journal} {\bibinfo  {journal} {Annual Review of Condensed Matter
  Physics}\ }\textbf {\bibinfo {volume} {1}},\ \bibinfo {pages} {179} (\bibinfo
  {year} {2010})}\BibitemShut {NoStop}%
\bibitem [{\citenamefont {Reimers}\ \emph {et~al.}(1991)\citenamefont
  {Reimers}, \citenamefont {Berlinsky},\ and\ \citenamefont
  {Shi}}]{reimers_1991}%
  \BibitemOpen
  \bibfield  {author} {\bibinfo {author} {\bibfnamefont {J.~N.}\ \bibnamefont
  {Reimers}}, \bibinfo {author} {\bibfnamefont {A.~J.}\ \bibnamefont
  {Berlinsky}}, \ and\ \bibinfo {author} {\bibfnamefont {A.~C.}\ \bibnamefont
  {Shi}},\ }\href {http://link.aps.org/doi/10.1103/PhysRevB.43.865} {\bibfield
  {journal} {\bibinfo  {journal} {Physical Review B}\ }\textbf {\bibinfo
  {volume} {43}},\ \bibinfo {pages} {865} (\bibinfo {year} {1991})}\BibitemShut
  {NoStop}%
\bibitem [{\citenamefont {Tchernyshyov}\ \emph {et~al.}(2002)\citenamefont
  {Tchernyshyov}, \citenamefont {Moessner},\ and\ \citenamefont
  {Sondhi}}]{tchernyshyov_2002}%
  \BibitemOpen
  \bibfield  {author} {\bibinfo {author} {\bibfnamefont {O.}~\bibnamefont
  {Tchernyshyov}}, \bibinfo {author} {\bibfnamefont {R.}~\bibnamefont
  {Moessner}}, \ and\ \bibinfo {author} {\bibfnamefont {S.~L.}\ \bibnamefont
  {Sondhi}},\ }\href {http://link.aps.org/doi/10.1103/PhysRevLett.88.067203}
  {\bibfield  {journal} {\bibinfo  {journal} {Physical Review Letters}\
  }\textbf {\bibinfo {volume} {88}},\ \bibinfo {pages} {067203} (\bibinfo
  {year} {2002})}\BibitemShut {NoStop}%
\bibitem [{\citenamefont {Jensen}\ and\ \citenamefont
  {Mackintosh}(1991)}]{Jensen:1991ux}%
  \BibitemOpen
  \bibfield  {author} {\bibinfo {author} {\bibfnamefont {J.}~\bibnamefont
  {Jensen}}\ and\ \bibinfo {author} {\bibfnamefont {A.~R.}\ \bibnamefont
  {Mackintosh}},\ }\href@noop {} {\emph {\bibinfo {title} {Rare Earth
  Magnetism}}}\ (\bibinfo  {publisher} {Clarendon Press, Oxford},\ \bibinfo
  {year} {1991})\BibitemShut {NoStop}%
\bibitem [{\citenamefont {Cheong}\ and\ \citenamefont
  {Mostovoy}(2007)}]{cheong_2007}%
  \BibitemOpen
  \bibfield  {author} {\bibinfo {author} {\bibfnamefont {S.-W.}\ \bibnamefont
  {Cheong}}\ and\ \bibinfo {author} {\bibfnamefont {M.}~\bibnamefont
  {Mostovoy}},\ }\href {\doibase 10.1038/nmat1804} {\bibfield  {journal}
  {\bibinfo  {journal} {Nature Materials}\ }\textbf {\bibinfo {volume} {6}},\
  \bibinfo {pages} {13} (\bibinfo {year} {2007})}\BibitemShut {NoStop}%
\bibitem [{\citenamefont {Mostovoy}(2006)}]{mostovoy_2006}%
  \BibitemOpen
  \bibfield  {author} {\bibinfo {author} {\bibfnamefont {M.}~\bibnamefont
  {Mostovoy}},\ }\href {http://link.aps.org/doi/10.1103/PhysRevLett.96.067601}
  {\bibfield  {journal} {\bibinfo  {journal} {Physical Review Letters}\
  }\textbf {\bibinfo {volume} {96}},\ \bibinfo {pages} {067601} (\bibinfo
  {year} {2006})}\BibitemShut {NoStop}%
\bibitem [{\citenamefont {Nagaosa}\ and\ \citenamefont
  {Tokura}(2013)}]{nagaosa_2013}%
  \BibitemOpen
  \bibfield  {author} {\bibinfo {author} {\bibfnamefont {N.}~\bibnamefont
  {Nagaosa}}\ and\ \bibinfo {author} {\bibfnamefont {Y.}~\bibnamefont
  {Tokura}},\ }\href {http://dx.doi.org/10.1038/nnano.2013.243} {\bibfield
  {journal} {\bibinfo  {journal} {Nature Nanotechnology}\ }\textbf {\bibinfo
  {volume} {8}},\ \bibinfo {pages} {899} (\bibinfo {year} {2013})}\BibitemShut
  {NoStop}%
\bibitem [{\citenamefont {M{\"u}hlbauer}\ \emph {et~al.}(2009)\citenamefont
  {M{\"u}hlbauer}, \citenamefont {Binz}, \citenamefont {Jonietz}, \citenamefont
  {Pfleiderer}, \citenamefont {Rosch}, \citenamefont {Neubauer}, \citenamefont
  {Georgii},\ and\ \citenamefont {B{\"o}ni}}]{muhlbauer_2009}%
  \BibitemOpen
  \bibfield  {author} {\bibinfo {author} {\bibfnamefont {S.}~\bibnamefont
  {M{\"u}hlbauer}}, \bibinfo {author} {\bibfnamefont {B.}~\bibnamefont {Binz}},
  \bibinfo {author} {\bibfnamefont {F.}~\bibnamefont {Jonietz}}, \bibinfo
  {author} {\bibfnamefont {C.}~\bibnamefont {Pfleiderer}}, \bibinfo {author}
  {\bibfnamefont {A.}~\bibnamefont {Rosch}}, \bibinfo {author} {\bibfnamefont
  {A.}~\bibnamefont {Neubauer}}, \bibinfo {author} {\bibfnamefont
  {R.}~\bibnamefont {Georgii}}, \ and\ \bibinfo {author} {\bibfnamefont
  {P.}~\bibnamefont {B{\"o}ni}},\ }\href
  {http://www.sciencemag.org/content/323/5916/915.abstract} {\bibfield
  {journal} {\bibinfo  {journal} {Science}\ }\textbf {\bibinfo {volume}
  {323}},\ \bibinfo {pages} {915} (\bibinfo {year} {2009})}\BibitemShut
  {NoStop}%
\bibitem [{\citenamefont {Bergman}\ \emph {et~al.}(2007)\citenamefont
  {Bergman}, \citenamefont {Alicea}, \citenamefont {Gull}, \citenamefont
  {Trebst},\ and\ \citenamefont {Balents}}]{bergman_2007}%
  \BibitemOpen
  \bibfield  {author} {\bibinfo {author} {\bibfnamefont {D.}~\bibnamefont
  {Bergman}}, \bibinfo {author} {\bibfnamefont {J.}~\bibnamefont {Alicea}},
  \bibinfo {author} {\bibfnamefont {E.}~\bibnamefont {Gull}}, \bibinfo {author}
  {\bibfnamefont {S.}~\bibnamefont {Trebst}}, \ and\ \bibinfo {author}
  {\bibfnamefont {L.}~\bibnamefont {Balents}},\ }\href {\doibase
  http://www.nature.com/nphys/journal/v3/n7/suppinfo/nphys622_S1.html}
  {\bibfield  {journal} {\bibinfo  {journal} {Nature Physics}\ }\textbf
  {\bibinfo {volume} {3}},\ \bibinfo {pages} {487} (\bibinfo {year}
  {2007})}\BibitemShut {NoStop}%
\bibitem [{\citenamefont {Lee}\ and\ \citenamefont {Balents}(2008)}]{lee_2008}%
  \BibitemOpen
  \bibfield  {author} {\bibinfo {author} {\bibfnamefont {S.~B.}\ \bibnamefont
  {Lee}}\ and\ \bibinfo {author} {\bibfnamefont {L.}~\bibnamefont {Balents}},\
  }\href {http://link.aps.org/doi/10.1103/PhysRevB.78.144417} {\bibfield
  {journal} {\bibinfo  {journal} {Physical Review B}\ }\textbf {\bibinfo
  {volume} {78}},\ \bibinfo {pages} {144417} (\bibinfo {year}
  {2008})}\BibitemShut {NoStop}%
\bibitem [{\citenamefont {Savary}\ \emph {et~al.}(2011)\citenamefont {Savary},
  \citenamefont {Gull}, \citenamefont {Trebst}, \citenamefont {Alicea},
  \citenamefont {Bergman},\ and\ \citenamefont {Balents}}]{savary_2011}%
  \BibitemOpen
  \bibfield  {author} {\bibinfo {author} {\bibfnamefont {L.}~\bibnamefont
  {Savary}}, \bibinfo {author} {\bibfnamefont {E.}~\bibnamefont {Gull}},
  \bibinfo {author} {\bibfnamefont {S.}~\bibnamefont {Trebst}}, \bibinfo
  {author} {\bibfnamefont {J.}~\bibnamefont {Alicea}}, \bibinfo {author}
  {\bibfnamefont {D.}~\bibnamefont {Bergman}}, \ and\ \bibinfo {author}
  {\bibfnamefont {L.}~\bibnamefont {Balents}},\ }\href
  {http://link.aps.org/doi/10.1103/PhysRevB.84.064438} {\bibfield  {journal}
  {\bibinfo  {journal} {Physical Review B}\ }\textbf {\bibinfo {volume} {84}},\
  \bibinfo {pages} {064438} (\bibinfo {year} {2011})}\BibitemShut {NoStop}%
\bibitem [{\citenamefont {Villain}\ \emph {et~al.}(1980)\citenamefont
  {Villain}, \citenamefont {Bidaux}, \citenamefont {Carton},\ and\
  \citenamefont {Conte}}]{villain_1980}%
  \BibitemOpen
  \bibfield  {author} {\bibinfo {author} {\bibfnamefont {J.}~\bibnamefont
  {Villain}}, \bibinfo {author} {\bibfnamefont {R.}~\bibnamefont {Bidaux}},
  \bibinfo {author} {\bibfnamefont {J.-P.}\ \bibnamefont {Carton}}, \ and\
  \bibinfo {author} {\bibfnamefont {R.}~\bibnamefont {Conte}},\ }\href
  {http://jphys.journaldephysique.org/articles/jphys/abs/1980/11/jphys_1980__41_11_1263_0/jphys_1980__41_11_1263_0.html}
  {\bibfield  {journal} {\bibinfo  {journal} {Journal de Physique}\ }\textbf
  {\bibinfo {volume} {41}},\ \bibinfo {pages} {1263} (\bibinfo {year}
  {1980})}\BibitemShut {NoStop}%
\bibitem [{\citenamefont {Henley}(1989)}]{henley_1989}%
  \BibitemOpen
  \bibfield  {author} {\bibinfo {author} {\bibfnamefont {C.~L.}\ \bibnamefont
  {Henley}},\ }\href {http://link.aps.org/doi/10.1103/PhysRevLett.62.2056}
  {\bibfield  {journal} {\bibinfo  {journal} {Physical Review Letters}\
  }\textbf {\bibinfo {volume} {62}},\ \bibinfo {pages} {2056} (\bibinfo {year}
  {1989})}\BibitemShut {NoStop}%
\bibitem [{\citenamefont {{T. Suzuki and H. Nagai and M. Nohara and H.
  Takagi}}(2007)}]{suzuki_2007}%
  \BibitemOpen
  \bibfield  {author} {\bibinfo {author} {\bibnamefont {{T. Suzuki and H. Nagai
  and M. Nohara and H. Takagi}}},\ }\href
  {http://stacks.iop.org/0953-8984/19/i=14/a=145265} {\bibfield  {journal}
  {\bibinfo  {journal} {Journal of Physics: Condensed Matter}\ }\textbf
  {\bibinfo {volume} {19}},\ \bibinfo {pages} {145265} (\bibinfo {year}
  {2007})}\BibitemShut {NoStop}%
\bibitem [{\citenamefont {Tristan}\ \emph {et~al.}(2005)\citenamefont
  {Tristan}, \citenamefont {Hemberger}, \citenamefont {Krimmel}, \citenamefont
  {Krug~von Nidda}, \citenamefont {Tsurkan},\ and\ \citenamefont
  {Loidl}}]{tristan_2005}%
  \BibitemOpen
  \bibfield  {author} {\bibinfo {author} {\bibfnamefont {N.}~\bibnamefont
  {Tristan}}, \bibinfo {author} {\bibfnamefont {J.}~\bibnamefont {Hemberger}},
  \bibinfo {author} {\bibfnamefont {A.}~\bibnamefont {Krimmel}}, \bibinfo
  {author} {\bibfnamefont {H.-A.}\ \bibnamefont {Krug~von Nidda}}, \bibinfo
  {author} {\bibfnamefont {V.}~\bibnamefont {Tsurkan}}, \ and\ \bibinfo
  {author} {\bibfnamefont {A.}~\bibnamefont {Loidl}},\ }\href
  {http://link.aps.org/doi/10.1103/PhysRevB.72.174404} {\bibfield  {journal}
  {\bibinfo  {journal} {Physical Review B}\ }\textbf {\bibinfo {volume} {72}},\
  \bibinfo {pages} {174404} (\bibinfo {year} {2005})}\BibitemShut {NoStop}%
\bibitem [{\citenamefont {Krimmel}\ \emph {et~al.}(2009)\citenamefont
  {Krimmel}, \citenamefont {Mutka}, \citenamefont {Koza}, \citenamefont
  {Tsurkan},\ and\ \citenamefont {Loidl}}]{krimmel_2009}%
  \BibitemOpen
  \bibfield  {author} {\bibinfo {author} {\bibfnamefont {A.}~\bibnamefont
  {Krimmel}}, \bibinfo {author} {\bibfnamefont {H.}~\bibnamefont {Mutka}},
  \bibinfo {author} {\bibfnamefont {M.~M.}\ \bibnamefont {Koza}}, \bibinfo
  {author} {\bibfnamefont {V.}~\bibnamefont {Tsurkan}}, \ and\ \bibinfo
  {author} {\bibfnamefont {A.}~\bibnamefont {Loidl}},\ }\href
  {http://link.aps.org/doi/10.1103/PhysRevB.79.134406} {\bibfield  {journal}
  {\bibinfo  {journal} {Physical Review B}\ }\textbf {\bibinfo {volume} {79}},\
  \bibinfo {pages} {134406} (\bibinfo {year} {2009})}\BibitemShut {NoStop}%
\bibitem [{\citenamefont {Zaharko}\ \emph {et~al.}(2011)\citenamefont
  {Zaharko}, \citenamefont {Christensen}, \citenamefont {Cervellino},
  \citenamefont {Tsurkan}, \citenamefont {Maljuk}, \citenamefont {Stuhr},
  \citenamefont {Niedermayer}, \citenamefont {Yokaichiya}, \citenamefont
  {Argyriou}, \citenamefont {Boehm},\ and\ \citenamefont
  {Loidl}}]{zaharko_2011}%
  \BibitemOpen
  \bibfield  {author} {\bibinfo {author} {\bibfnamefont {O.}~\bibnamefont
  {Zaharko}}, \bibinfo {author} {\bibfnamefont {N.~B.}\ \bibnamefont
  {Christensen}}, \bibinfo {author} {\bibfnamefont {A.}~\bibnamefont
  {Cervellino}}, \bibinfo {author} {\bibfnamefont {V.}~\bibnamefont {Tsurkan}},
  \bibinfo {author} {\bibfnamefont {A.}~\bibnamefont {Maljuk}}, \bibinfo
  {author} {\bibfnamefont {U.}~\bibnamefont {Stuhr}}, \bibinfo {author}
  {\bibfnamefont {C.}~\bibnamefont {Niedermayer}}, \bibinfo {author}
  {\bibfnamefont {F.}~\bibnamefont {Yokaichiya}}, \bibinfo {author}
  {\bibfnamefont {D.~N.}\ \bibnamefont {Argyriou}}, \bibinfo {author}
  {\bibfnamefont {M.}~\bibnamefont {Boehm}}, \ and\ \bibinfo {author}
  {\bibfnamefont {A.}~\bibnamefont {Loidl}},\ }\href
  {http://link.aps.org/doi/10.1103/PhysRevB.84.094403} {\bibfield  {journal}
  {\bibinfo  {journal} {Physical Review B}\ }\textbf {\bibinfo {volume} {84}},\
  \bibinfo {pages} {094403} (\bibinfo {year} {2011})}\BibitemShut {NoStop}%
\bibitem [{\citenamefont {MacDougall}\ \emph {et~al.}(2011)\citenamefont
  {MacDougall}, \citenamefont {Gout}, \citenamefont {Zarestky}, \citenamefont
  {Ehlers}, \citenamefont {Podlesnyak}, \citenamefont {McGuire}, \citenamefont
  {Mandrus},\ and\ \citenamefont {Nagler}}]{macdougall_2011}%
  \BibitemOpen
  \bibfield  {author} {\bibinfo {author} {\bibfnamefont {G.~J.}\ \bibnamefont
  {MacDougall}}, \bibinfo {author} {\bibfnamefont {D.}~\bibnamefont {Gout}},
  \bibinfo {author} {\bibfnamefont {J.~L.}\ \bibnamefont {Zarestky}}, \bibinfo
  {author} {\bibfnamefont {G.}~\bibnamefont {Ehlers}}, \bibinfo {author}
  {\bibfnamefont {A.}~\bibnamefont {Podlesnyak}}, \bibinfo {author}
  {\bibfnamefont {M.~A.}\ \bibnamefont {McGuire}}, \bibinfo {author}
  {\bibfnamefont {D.}~\bibnamefont {Mandrus}}, \ and\ \bibinfo {author}
  {\bibfnamefont {S.~E.}\ \bibnamefont {Nagler}},\ }\href {\doibase
  10.1073/pnas.1107861108} {\bibfield  {journal} {\bibinfo  {journal}
  {Proceedings of the National Academy of Sciences}\ }\textbf {\bibinfo
  {volume} {108}},\ \bibinfo {pages} {15693} (\bibinfo {year}
  {2011})}\BibitemShut {NoStop}%
\bibitem [{\citenamefont {Fritsch}\ \emph {et~al.}(2004)\citenamefont
  {Fritsch}, \citenamefont {Hemberger}, \citenamefont {B{\"u}ttgen},
  \citenamefont {Scheidt}, \citenamefont {Krug~von Nidda}, \citenamefont
  {Loidl},\ and\ \citenamefont {Tsurkan}}]{fritsch_2004}%
  \BibitemOpen
  \bibfield  {author} {\bibinfo {author} {\bibfnamefont {V.}~\bibnamefont
  {Fritsch}}, \bibinfo {author} {\bibfnamefont {J.}~\bibnamefont {Hemberger}},
  \bibinfo {author} {\bibfnamefont {N.}~\bibnamefont {B{\"u}ttgen}}, \bibinfo
  {author} {\bibfnamefont {E.~W.}\ \bibnamefont {Scheidt}}, \bibinfo {author}
  {\bibfnamefont {H.~A.}\ \bibnamefont {Krug~von Nidda}}, \bibinfo {author}
  {\bibfnamefont {A.}~\bibnamefont {Loidl}}, \ and\ \bibinfo {author}
  {\bibfnamefont {V.}~\bibnamefont {Tsurkan}},\ }\href
  {http://link.aps.org/doi/10.1103/PhysRevLett.92.116401} {\bibfield  {journal}
  {\bibinfo  {journal} {Physical Review Letters}\ }\textbf {\bibinfo {volume}
  {92}},\ \bibinfo {pages} {116401} (\bibinfo {year} {2004})}\BibitemShut
  {NoStop}%
\bibitem [{\citenamefont {Chen}\ \emph {et~al.}(2009)\citenamefont {Chen},
  \citenamefont {Balents},\ and\ \citenamefont {Schnyder}}]{chen_2009}%
  \BibitemOpen
  \bibfield  {author} {\bibinfo {author} {\bibfnamefont {G.}~\bibnamefont
  {Chen}}, \bibinfo {author} {\bibfnamefont {L.}~\bibnamefont {Balents}}, \
  and\ \bibinfo {author} {\bibfnamefont {A.~P.}\ \bibnamefont {Schnyder}},\
  }\href {http://link.aps.org/doi/10.1103/PhysRevLett.102.096406} {\bibfield
  {journal} {\bibinfo  {journal} {Physical Review Letters}\ }\textbf {\bibinfo
  {volume} {102}},\ \bibinfo {pages} {096406} (\bibinfo {year}
  {2009})}\BibitemShut {NoStop}%
\bibitem [{\citenamefont {Krimmel}\ \emph {et~al.}(2006)\citenamefont
  {Krimmel}, \citenamefont {M{\"u}cksch}, \citenamefont {Tsurkan},
  \citenamefont {Koza}, \citenamefont {Mutka}, \citenamefont {Ritter},
  \citenamefont {Sheptyakov}, \citenamefont {Horn},\ and\ \citenamefont
  {Loidl}}]{krimmel_2006}%
  \BibitemOpen
  \bibfield  {author} {\bibinfo {author} {\bibfnamefont {A.}~\bibnamefont
  {Krimmel}}, \bibinfo {author} {\bibfnamefont {M.}~\bibnamefont
  {M{\"u}cksch}}, \bibinfo {author} {\bibfnamefont {V.}~\bibnamefont
  {Tsurkan}}, \bibinfo {author} {\bibfnamefont {M.~M.}\ \bibnamefont {Koza}},
  \bibinfo {author} {\bibfnamefont {H.}~\bibnamefont {Mutka}}, \bibinfo
  {author} {\bibfnamefont {C.}~\bibnamefont {Ritter}}, \bibinfo {author}
  {\bibfnamefont {D.~V.}\ \bibnamefont {Sheptyakov}}, \bibinfo {author}
  {\bibfnamefont {S.}~\bibnamefont {Horn}}, \ and\ \bibinfo {author}
  {\bibfnamefont {A.}~\bibnamefont {Loidl}},\ }\href
  {http://link.aps.org/doi/10.1103/PhysRevB.73.014413} {\bibfield  {journal}
  {\bibinfo  {journal} {Physical Review B}\ }\textbf {\bibinfo {volume} {73}},\
  \bibinfo {pages} {014413} (\bibinfo {year} {2006})}\BibitemShut {NoStop}%
\bibitem [{\citenamefont {{M. M{\"u}cksch and M. M. Koza and H. Mutka and C.
  Ritter and A. Cervellino and A. Podlesnyak and D. Sheptyakov and V. Tsurkan
  and A. Krimmel and S. Horn and A. Loidl}}(2007)}]{mucksch_2007}%
  \BibitemOpen
  \bibfield  {author} {\bibinfo {author} {\bibnamefont {{M. M{\"u}cksch and M.
  M. Koza and H. Mutka and C. Ritter and A. Cervellino and A. Podlesnyak and D.
  Sheptyakov and V. Tsurkan and A. Krimmel and S. Horn and A. Loidl}}},\ }\href
  {http://stacks.iop.org/0953-8984/19/i=14/a=145262} {\bibfield  {journal}
  {\bibinfo  {journal} {Journal of Physics: Condensed Matter}\ }\textbf
  {\bibinfo {volume} {19}},\ \bibinfo {pages} {145262} (\bibinfo {year}
  {2007})}\BibitemShut {NoStop}%
\bibitem [{\citenamefont {Giri}\ \emph {et~al.}(2005)\citenamefont {Giri},
  \citenamefont {Nakamura},\ and\ \citenamefont {Kohara}}]{giri_2005}%
  \BibitemOpen
  \bibfield  {author} {\bibinfo {author} {\bibfnamefont {S.}~\bibnamefont
  {Giri}}, \bibinfo {author} {\bibfnamefont {H.}~\bibnamefont {Nakamura}}, \
  and\ \bibinfo {author} {\bibfnamefont {T.}~\bibnamefont {Kohara}},\ }\href
  {http://link.aps.org/doi/10.1103/PhysRevB.72.132404} {\bibfield  {journal}
  {\bibinfo  {journal} {Physical Review B}\ }\textbf {\bibinfo {volume} {72}},\
  \bibinfo {pages} {132404} (\bibinfo {year} {2005})}\BibitemShut {NoStop}%
\bibitem [{\citenamefont {B{\"u}ttgen}\ \emph {et~al.}(2006)\citenamefont
  {B{\"u}ttgen}, \citenamefont {Zymara}, \citenamefont {Kegler}, \citenamefont
  {Tsurkan},\ and\ \citenamefont {Loidl}}]{buttgen_2006}%
  \BibitemOpen
  \bibfield  {author} {\bibinfo {author} {\bibfnamefont {N.}~\bibnamefont
  {B{\"u}ttgen}}, \bibinfo {author} {\bibfnamefont {A.}~\bibnamefont {Zymara}},
  \bibinfo {author} {\bibfnamefont {C.}~\bibnamefont {Kegler}}, \bibinfo
  {author} {\bibfnamefont {V.}~\bibnamefont {Tsurkan}}, \ and\ \bibinfo
  {author} {\bibfnamefont {A.}~\bibnamefont {Loidl}},\ }\href
  {http://link.aps.org/doi/10.1103/PhysRevB.73.132409} {\bibfield  {journal}
  {\bibinfo  {journal} {Physical Review B}\ }\textbf {\bibinfo {volume} {73}},\
  \bibinfo {pages} {132409} (\bibinfo {year} {2006})}\BibitemShut {NoStop}%
\bibitem [{\citenamefont {Kalvius}\ \emph {et~al.}(2006)\citenamefont
  {Kalvius}, \citenamefont {Hartmann}, \citenamefont {Noakes}, \citenamefont
  {Wagner}, \citenamefont {W{\"a}ppling}, \citenamefont {Zimmermann},
  \citenamefont {Baines}, \citenamefont {Krimmel}, \citenamefont {Tsurkan},\
  and\ \citenamefont {Loidl}}]{kalvius_2006}%
  \BibitemOpen
  \bibfield  {author} {\bibinfo {author} {\bibfnamefont {G.~M.}\ \bibnamefont
  {Kalvius}}, \bibinfo {author} {\bibfnamefont {O.}~\bibnamefont {Hartmann}},
  \bibinfo {author} {\bibfnamefont {D.~R.}\ \bibnamefont {Noakes}}, \bibinfo
  {author} {\bibfnamefont {F.~E.}\ \bibnamefont {Wagner}}, \bibinfo {author}
  {\bibfnamefont {R.}~\bibnamefont {W{\"a}ppling}}, \bibinfo {author}
  {\bibfnamefont {U.}~\bibnamefont {Zimmermann}}, \bibinfo {author}
  {\bibfnamefont {C.}~\bibnamefont {Baines}}, \bibinfo {author} {\bibfnamefont
  {A.}~\bibnamefont {Krimmel}}, \bibinfo {author} {\bibfnamefont
  {V.}~\bibnamefont {Tsurkan}}, \ and\ \bibinfo {author} {\bibfnamefont
  {A.}~\bibnamefont {Loidl}},\ }\href {\doibase 10.1016/j.physb.2006.01.158}
  {\bibfield  {journal} {\bibinfo  {journal} {Proceedings of the International
  Conference on Strongly Correlated Electron Systems SCES 2005}\ }\textbf
  {\bibinfo {volume} {378–-380}},\ \bibinfo {pages} {592} (\bibinfo {year}
  {2006})}\BibitemShut {NoStop}%
\bibitem [{\citenamefont {Okubo}\ \emph {et~al.}(2012)\citenamefont {Okubo},
  \citenamefont {Chung},\ and\ \citenamefont {Kawamura}}]{okubo_2012}%
  \BibitemOpen
  \bibfield  {author} {\bibinfo {author} {\bibfnamefont {T.}~\bibnamefont
  {Okubo}}, \bibinfo {author} {\bibfnamefont {S.}~\bibnamefont {Chung}}, \ and\
  \bibinfo {author} {\bibfnamefont {H.}~\bibnamefont {Kawamura}},\ }\href
  {http://link.aps.org/doi/10.1103/PhysRevLett.108.017206} {\bibfield
  {journal} {\bibinfo  {journal} {Physical Review Letters}\ }\textbf {\bibinfo
  {volume} {108}},\ \bibinfo {pages} {017206} (\bibinfo {year}
  {2012})}\BibitemShut {NoStop}%
\bibitem [{\citenamefont {Kamiya}\ and\ \citenamefont
  {Batista}(2014)}]{kamiya_2014}%
  \BibitemOpen
  \bibfield  {author} {\bibinfo {author} {\bibfnamefont {Y.}~\bibnamefont
  {Kamiya}}\ and\ \bibinfo {author} {\bibfnamefont {C.}~\bibnamefont
  {Batista}},\ }\href {http://link.aps.org/doi/10.1103/PhysRevX.4.011023}
  {\bibfield  {journal} {\bibinfo  {journal} {Physical Review X}\ }\textbf
  {\bibinfo {volume} {4}},\ \bibinfo {pages} {011023} (\bibinfo {year}
  {2014})}\BibitemShut {NoStop}%
\bibitem [{\citenamefont {Wang}\ \emph {et~al.}(2015)\citenamefont {Wang},
  \citenamefont {Kamiya}, \citenamefont {Nevidomskyy},\ and\ \citenamefont
  {Batista}}]{wang_2015}%
  \BibitemOpen
  \bibfield  {author} {\bibinfo {author} {\bibfnamefont {Z.}~\bibnamefont
  {Wang}}, \bibinfo {author} {\bibfnamefont {Y.}~\bibnamefont {Kamiya}},
  \bibinfo {author} {\bibfnamefont {A.~H.}\ \bibnamefont {Nevidomskyy}}, \ and\
  \bibinfo {author} {\bibfnamefont {C.~D.}\ \bibnamefont {Batista}},\ }\href
  {http://link.aps.org/doi/10.1103/PhysRevLett.115.107201} {\bibfield
  {journal} {\bibinfo  {journal} {Physical Review Letters}\ }\textbf {\bibinfo
  {volume} {115}},\ \bibinfo {pages} {107201} (\bibinfo {year}
  {2015})}\BibitemShut {NoStop}%
\bibitem [{\citenamefont {Rousochatzakis}\ \emph {et~al.}(2016)\citenamefont
  {Rousochatzakis}, \citenamefont {R{\"o}ssler}, \citenamefont {van~den
  Brink},\ and\ \citenamefont {Daghofer}}]{rousochatzakis_2016}%
  \BibitemOpen
  \bibfield  {author} {\bibinfo {author} {\bibfnamefont {I.}~\bibnamefont
  {Rousochatzakis}}, \bibinfo {author} {\bibfnamefont {U.~K.}\ \bibnamefont
  {R{\"o}ssler}}, \bibinfo {author} {\bibfnamefont {J.}~\bibnamefont {van~den
  Brink}}, \ and\ \bibinfo {author} {\bibfnamefont {M.}~\bibnamefont
  {Daghofer}},\ }\href {http://link.aps.org/doi/10.1103/PhysRevB.93.104417}
  {\bibfield  {journal} {\bibinfo  {journal} {Physical Review B}\ }\textbf
  {\bibinfo {volume} {93}},\ \bibinfo {pages} {104417} (\bibinfo {year}
  {2016})}\BibitemShut {NoStop}%
\bibitem [{\citenamefont {Reimers}(1992)}]{reimers_1992}%
  \BibitemOpen
  \bibfield  {author} {\bibinfo {author} {\bibfnamefont {J.~N.}\ \bibnamefont
  {Reimers}},\ }\href {http://link.aps.org/doi/10.1103/PhysRevB.46.193}
  {\bibfield  {journal} {\bibinfo  {journal} {Physical Review B}\ }\textbf
  {\bibinfo {volume} {46}},\ \bibinfo {pages} {193} (\bibinfo {year}
  {1992})}\BibitemShut {NoStop}%
\bibitem [{\citenamefont {Fennell}\ \emph {et~al.}(2009)\citenamefont
  {Fennell}, \citenamefont {Deen}, \citenamefont {Wildes}, \citenamefont
  {Schmalzl}, \citenamefont {Prabhakaran}, \citenamefont {Boothroyd},
  \citenamefont {Aldus}, \citenamefont {McMorrow},\ and\ \citenamefont
  {Bramwell}}]{Fennell_2009}%
  \BibitemOpen
  \bibfield  {author} {\bibinfo {author} {\bibfnamefont {T.}~\bibnamefont
  {Fennell}}, \bibinfo {author} {\bibfnamefont {P.~P.}\ \bibnamefont {Deen}},
  \bibinfo {author} {\bibfnamefont {A.~R.}\ \bibnamefont {Wildes}}, \bibinfo
  {author} {\bibfnamefont {K.}~\bibnamefont {Schmalzl}}, \bibinfo {author}
  {\bibfnamefont {D.}~\bibnamefont {Prabhakaran}}, \bibinfo {author}
  {\bibfnamefont {A.~T.}\ \bibnamefont {Boothroyd}}, \bibinfo {author}
  {\bibfnamefont {R.~J.}\ \bibnamefont {Aldus}}, \bibinfo {author}
  {\bibfnamefont {D.~F.}\ \bibnamefont {McMorrow}}, \ and\ \bibinfo {author}
  {\bibfnamefont {S.~T.}\ \bibnamefont {Bramwell}},\ }\href {\doibase
  10.1126/science.1177582} {\bibfield  {journal} {\bibinfo  {journal}
  {Science}\ }\textbf {\bibinfo {volume} {326}},\ \bibinfo {pages} {415}
  (\bibinfo {year} {2009})}\BibitemShut {NoStop}%
\bibitem [{\citenamefont {Brown}(2006)}]{brown_2006}%
  \BibitemOpen
  \bibfield  {author} {\bibinfo {author} {\bibfnamefont {P.~J.}\ \bibnamefont
  {Brown}},\ }in\ \href@noop {} {\emph {\bibinfo {booktitle} {Neutron
  Scattering from Magnetic Materials}}},\ \bibinfo {editor} {edited by\
  \bibinfo {editor} {\bibfnamefont {T.}~\bibnamefont {Chatterji}}}\ (\bibinfo
  {publisher} {Elsevier B. V.},\ \bibinfo {year} {2006})\BibitemShut {NoStop}%
\bibitem [{\citenamefont {Mochizuki}\ and\ \citenamefont
  {Furukawa}(2009)}]{mochizuki_2009}%
  \BibitemOpen
  \bibfield  {author} {\bibinfo {author} {\bibfnamefont {M.}~\bibnamefont
  {Mochizuki}}\ and\ \bibinfo {author} {\bibfnamefont {N.}~\bibnamefont
  {Furukawa}},\ }\href {http://link.aps.org/doi/10.1103/PhysRevB.80.134416}
  {\bibfield  {journal} {\bibinfo  {journal} {Physical Review B}\ }\textbf
  {\bibinfo {volume} {80}},\ \bibinfo {pages} {134416} (\bibinfo {year}
  {2009})}\BibitemShut {NoStop}%
\bibitem [{\citenamefont {Mulder}\ \emph {et~al.}(2010)\citenamefont {Mulder},
  \citenamefont {Ganesh}, \citenamefont {Capriotti},\ and\ \citenamefont
  {Paramekanti}}]{mulder_2010}%
  \BibitemOpen
  \bibfield  {author} {\bibinfo {author} {\bibfnamefont {A.}~\bibnamefont
  {Mulder}}, \bibinfo {author} {\bibfnamefont {R.}~\bibnamefont {Ganesh}},
  \bibinfo {author} {\bibfnamefont {L.}~\bibnamefont {Capriotti}}, \ and\
  \bibinfo {author} {\bibfnamefont {A.}~\bibnamefont {Paramekanti}},\ }\href
  {http://link.aps.org/doi/10.1103/PhysRevB.81.214419} {\bibfield  {journal}
  {\bibinfo  {journal} {Physical Review B}\ }\textbf {\bibinfo {volume} {81}},\
  \bibinfo {pages} {214419} (\bibinfo {year} {2010})}\BibitemShut {NoStop}%
\bibitem [{\citenamefont {Rodríguez-Carvajal}(1993)}]{rodriguez_1993}%
  \BibitemOpen
  \bibfield  {author} {\bibinfo {author} {\bibfnamefont {J.}~\bibnamefont
  {Rodríguez-Carvajal}},\ }\href {\doibase 10.1016/0921-4526(93)90108-I}
  {\bibfield  {journal} {\bibinfo  {journal} {Physica B: Condensed Matter}\
  }\textbf {\bibinfo {volume} {192}},\ \bibinfo {pages} {55} (\bibinfo {year}
  {1993})}\BibitemShut {NoStop}%
\bibitem [{\citenamefont {Bauer}\ \emph {et~al.}(2011)\citenamefont {Bauer},
  \citenamefont {Carr}, \citenamefont {Evertz}, \citenamefont {Feiguin},
  \citenamefont {Freire}, \citenamefont {Fuchs}, \citenamefont {Gamper},
  \citenamefont {Gukelberger}, \citenamefont {Gull}, \citenamefont {Guertler},
  \citenamefont {Hehn}, \citenamefont {Igarashi}, \citenamefont {Isakov},
  \citenamefont {Koop}, \citenamefont {Ma}, \citenamefont {Mates},
  \citenamefont {Matsuo}, \citenamefont {Parcollet}, \citenamefont
  {Paw{\l}owski}, \citenamefont {Picon}, \citenamefont {Pollet}, \citenamefont
  {Santos}, \citenamefont {Scarola}, \citenamefont {Schollw{\"o}ck},
  \citenamefont {Silva}, \citenamefont {Surer}, \citenamefont {Todo},
  \citenamefont {Trebst}, \citenamefont {Troyer}, \citenamefont {Wall},
  \citenamefont {Werner},\ and\ \citenamefont {Wessel}}]{bauer_2011}%
  \BibitemOpen
  \bibfield  {author} {\bibinfo {author} {\bibfnamefont {B.}~\bibnamefont
  {Bauer}}, \bibinfo {author} {\bibfnamefont {L.~D.}\ \bibnamefont {Carr}},
  \bibinfo {author} {\bibfnamefont {H.~G.}\ \bibnamefont {Evertz}}, \bibinfo
  {author} {\bibfnamefont {A.}~\bibnamefont {Feiguin}}, \bibinfo {author}
  {\bibfnamefont {J.}~\bibnamefont {Freire}}, \bibinfo {author} {\bibfnamefont
  {S.}~\bibnamefont {Fuchs}}, \bibinfo {author} {\bibfnamefont
  {L.}~\bibnamefont {Gamper}}, \bibinfo {author} {\bibfnamefont
  {J.}~\bibnamefont {Gukelberger}}, \bibinfo {author} {\bibfnamefont
  {E.}~\bibnamefont {Gull}}, \bibinfo {author} {\bibfnamefont {S.}~\bibnamefont
  {Guertler}}, \bibinfo {author} {\bibfnamefont {A.}~\bibnamefont {Hehn}},
  \bibinfo {author} {\bibfnamefont {R.}~\bibnamefont {Igarashi}}, \bibinfo
  {author} {\bibfnamefont {S.~V.}\ \bibnamefont {Isakov}}, \bibinfo {author}
  {\bibfnamefont {D.}~\bibnamefont {Koop}}, \bibinfo {author} {\bibfnamefont
  {P.~N.}\ \bibnamefont {Ma}}, \bibinfo {author} {\bibfnamefont
  {P.}~\bibnamefont {Mates}}, \bibinfo {author} {\bibfnamefont
  {H.}~\bibnamefont {Matsuo}}, \bibinfo {author} {\bibfnamefont
  {O.}~\bibnamefont {Parcollet}}, \bibinfo {author} {\bibfnamefont
  {G.}~\bibnamefont {Paw{\l}owski}}, \bibinfo {author} {\bibfnamefont {J.~D.}\
  \bibnamefont {Picon}}, \bibinfo {author} {\bibfnamefont {L.}~\bibnamefont
  {Pollet}}, \bibinfo {author} {\bibfnamefont {E.}~\bibnamefont {Santos}},
  \bibinfo {author} {\bibfnamefont {V.~W.}\ \bibnamefont {Scarola}}, \bibinfo
  {author} {\bibfnamefont {U.}~\bibnamefont {Schollw{\"o}ck}}, \bibinfo
  {author} {\bibfnamefont {C.}~\bibnamefont {Silva}}, \bibinfo {author}
  {\bibfnamefont {B.}~\bibnamefont {Surer}}, \bibinfo {author} {\bibfnamefont
  {S.}~\bibnamefont {Todo}}, \bibinfo {author} {\bibfnamefont {S.}~\bibnamefont
  {Trebst}}, \bibinfo {author} {\bibfnamefont {M.}~\bibnamefont {Troyer}},
  \bibinfo {author} {\bibfnamefont {M.~L.}\ \bibnamefont {Wall}}, \bibinfo
  {author} {\bibfnamefont {P.}~\bibnamefont {Werner}}, \ and\ \bibinfo {author}
  {\bibfnamefont {S.}~\bibnamefont {Wessel}},\ }\href
  {http://stacks.iop.org/1742-5468/2011/i=05/a=P05001} {\bibfield  {journal}
  {\bibinfo  {journal} {Journal of Statistical Mechanics: Theory and
  Experiment}\ }\textbf {\bibinfo {volume} {2011}},\ \bibinfo {pages} {P05001}
  (\bibinfo {year} {2011})}\BibitemShut {NoStop}%
\end{thebibliography}%


\begin{thebibliography}{5}%
\makeatletter
\providecommand \@ifxundefined [1]{%
 \@ifx{#1\undefined}
}%
\providecommand \@ifnum [1]{%
 \ifnum #1\expandafter \@firstoftwo
 \else \expandafter \@secondoftwo
 \fi
}%
\providecommand \@ifx [1]{%
 \ifx #1\expandafter \@firstoftwo
 \else \expandafter \@secondoftwo
 \fi
}%
\providecommand \natexlab [1]{#1}%
\providecommand \enquote  [1]{``#1''}%
\providecommand \bibnamefont  [1]{#1}%
\providecommand \bibfnamefont [1]{#1}%
\providecommand \citenamefont [1]{#1}%
\providecommand \href@noop [0]{\@secondoftwo}%
\providecommand \href [0]{\begingroup \@sanitize@url \@href}%
\providecommand \@href[1]{\@@startlink{#1}\@@href}%
\providecommand \@@href[1]{\endgroup#1\@@endlink}%
\providecommand \@sanitize@url [0]{\catcode `\\12\catcode `\$12\catcode
  `\&12\catcode `\#12\catcode `\^12\catcode `\_12\catcode `\%12\relax}%
\providecommand \@@startlink[1]{}%
\providecommand \@@endlink[0]{}%
\providecommand \url  [0]{\begingroup\@sanitize@url \@url }%
\providecommand \@url [1]{\endgroup\@href {#1}{\urlprefix }}%
\providecommand \urlprefix  [0]{URL }%
\providecommand \Eprint [0]{\href }%
\providecommand \doibase [0]{http://dx.doi.org/}%
\providecommand \selectlanguage [0]{\@gobble}%
\providecommand \bibinfo  [0]{\@secondoftwo}%
\providecommand \bibfield  [0]{\@secondoftwo}%
\providecommand \translation [1]{[#1]}%
\providecommand \BibitemOpen [0]{}%
\providecommand \bibitemStop [0]{}%
\providecommand \bibitemNoStop [0]{.\EOS\space}%
\providecommand \EOS [0]{\spacefactor3000\relax}%
\providecommand \BibitemShut  [1]{\csname bibitem#1\endcsname}%
\let\auto@bib@innerbib\@empty
\bibitem [{\citenamefont {Rodriguez-Carvajal}(1993)}]{rodriguez_1993s}%
  \BibitemOpen
  \bibfield  {author} {\bibinfo {author} {\bibfnamefont {J.}~\bibnamefont
  {Rodriguez-Carvajal}},\ }\href@noop {} {\bibfield  {journal} {\bibinfo
  {journal} {Physica B: Condensed Matter}\ }\textbf {\bibinfo {volume} {192}},\
  \bibinfo {pages} {55} (\bibinfo {year} {1993})}\BibitemShut {NoStop}%
\bibitem [{\citenamefont {Vindigni}\ \emph {et~al.}(2008)\citenamefont
  {Vindigni}, \citenamefont {Saratz}, \citenamefont {Portmann}, \citenamefont
  {Pescia},\ and\ \citenamefont {Politi}}]{vindigni_2008}%
  \BibitemOpen
  \bibfield  {author} {\bibinfo {author} {\bibfnamefont {A.}~\bibnamefont
  {Vindigni}}, \bibinfo {author} {\bibfnamefont {N.}~\bibnamefont {Saratz}},
  \bibinfo {author} {\bibfnamefont {O.}~\bibnamefont {Portmann}}, \bibinfo
  {author} {\bibfnamefont {D.}~\bibnamefont {Pescia}}, \ and\ \bibinfo {author}
  {\bibfnamefont {P.}~\bibnamefont {Politi}},\ }\href {\doibase
  10.1103/PhysRevB.77.092414} {\bibfield  {journal} {\bibinfo  {journal}
  {Physical Review B}\ }\textbf {\bibinfo {volume} {77}},\ \bibinfo {pages}
  {092414} (\bibinfo {year} {2008})}\BibitemShut {NoStop}%
\bibitem [{\citenamefont {Pregelj}\ \emph {et~al.}(2015)\citenamefont
  {Pregelj}, \citenamefont {Zorko}, \citenamefont {Zaharko}, \citenamefont
  {Nojiri}, \citenamefont {Berger}, \citenamefont {Chapon},\ and\ \citenamefont
  {Arcon}}]{pregelj_2015s}%
  \BibitemOpen
  \bibfield  {author} {\bibinfo {author} {\bibfnamefont {M.}~\bibnamefont
  {Pregelj}}, \bibinfo {author} {\bibfnamefont {A.}~\bibnamefont {Zorko}},
  \bibinfo {author} {\bibfnamefont {O.}~\bibnamefont {Zaharko}}, \bibinfo
  {author} {\bibfnamefont {H.}~\bibnamefont {Nojiri}}, \bibinfo {author}
  {\bibfnamefont {H.}~\bibnamefont {Berger}}, \bibinfo {author} {\bibfnamefont
  {L.~C.}\ \bibnamefont {Chapon}}, \ and\ \bibinfo {author} {\bibfnamefont
  {D.}~\bibnamefont {Arcon}},\ }\href {http://dx.doi.org/10.1038/ncomms8255}
  {\bibfield  {journal} {\bibinfo  {journal} {Nature Communications}\ }\textbf
  {\bibinfo {volume} {6}},\ \bibinfo {pages} {7255} (\bibinfo {year}
  {2015})}\BibitemShut {NoStop}%
\bibitem [{\citenamefont {Lee}\ and\ \citenamefont {Balents}(2008)}]{lee_2008s}%
  \BibitemOpen
  \bibfield  {author} {\bibinfo {author} {\bibfnamefont {S.~B.}\ \bibnamefont
  {Lee}}\ and\ \bibinfo {author} {\bibfnamefont {L.}~\bibnamefont {Balents}},\
  }\href {http://link.aps.org/doi/10.1103/PhysRevB.78.144417} {\bibfield
  {journal} {\bibinfo  {journal} {Physical Review B}\ }\textbf {\bibinfo
  {volume} {78}},\ \bibinfo {pages} {144417} (\bibinfo {year}
  {2008})}\BibitemShut {NoStop}%
\bibitem [{\citenamefont {Nagaosa}\ and\ \citenamefont
  {Tokura}(2013)}]{nagaosa_2013s}%
  \BibitemOpen
  \bibfield  {author} {\bibinfo {author} {\bibfnamefont {N.}~\bibnamefont
  {Nagaosa}}\ and\ \bibinfo {author} {\bibfnamefont {Y.}~\bibnamefont
  {Tokura}},\ }\href {http://dx.doi.org/10.1038/nnano.2013.243} {\bibfield
  {journal} {\bibinfo  {journal} {Nature Nanotechnology}\ }\textbf {\bibinfo
  {volume} {8}},\ \bibinfo {pages} {899} (\bibinfo {year} {2013})}\BibitemShut
  {NoStop}%
\end{thebibliography}

\newpage

\renewcommand{\thefigure}{S\arabic{figure}}
\renewcommand{\thetable}{S\arabic{table}}

\renewcommand{\theequation}{\alph{equation}}

\makeatletter
\renewcommand{\@biblabel}[1]{\quad S#1. }
\makeatother

 \setcounter{figure}{0} 

\onecolumngrid
\begin{center} {\bf \large Spiral spin-liquid and the emergence of a vortex-like state in MnSc$_2$S$_4$ \\
 Supplementary Information} \end{center}
\vspace{0.5cm}
\twocolumngrid

\maketitle

\textbf{Synchrotron X-ray diffraction results}

\begin{figure} [h!]
\includegraphics[width=0.25\textwidth]{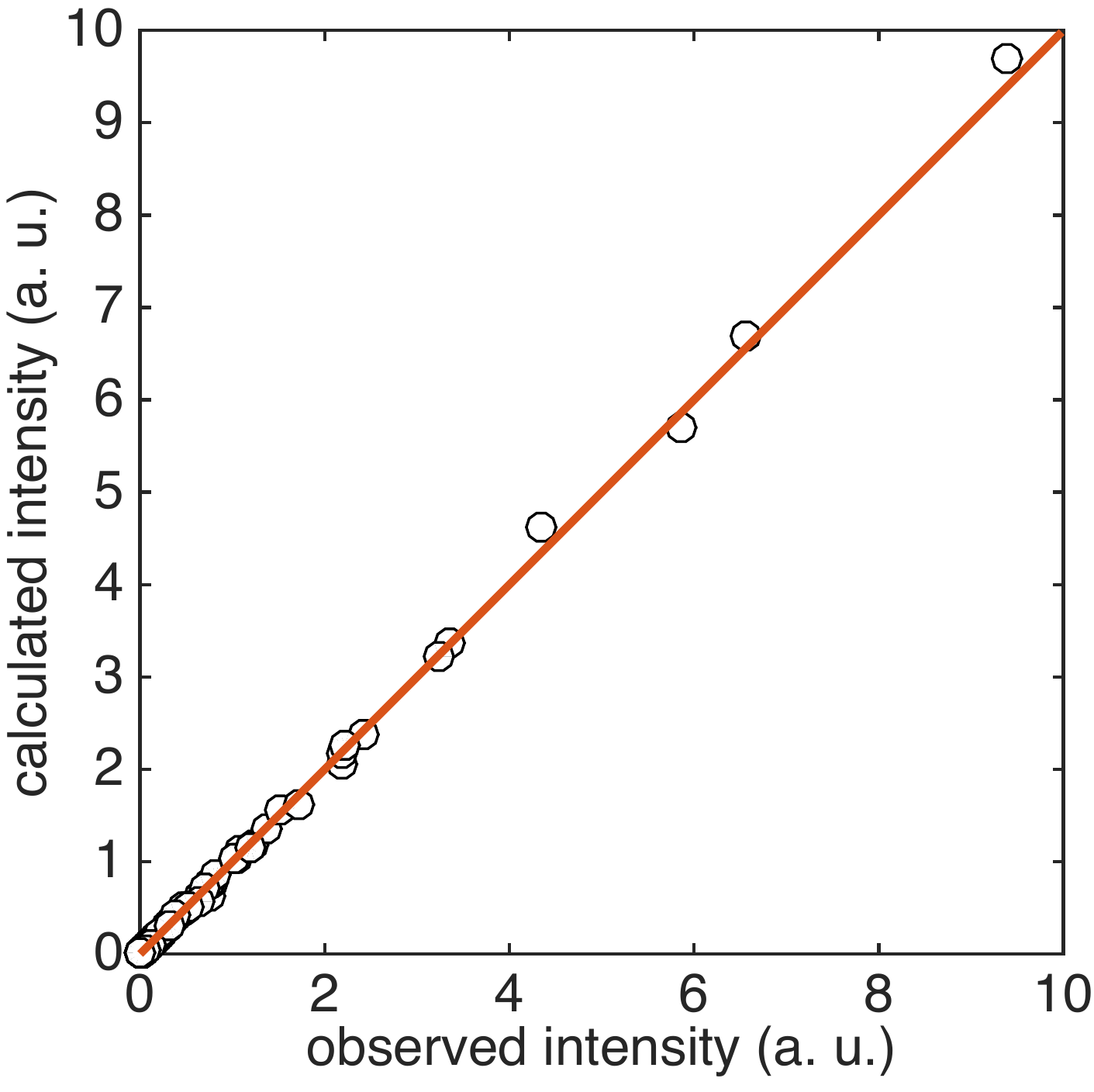}
\caption{\textbf{Refinement of the synchrotron X-ray diffraction data.} Comparison between the observed and calculated intensities. \label{fig:refinement}} 
\end{figure}

At room temperature, altogether 2738 Bragg reflections were collected on the Swiss-Norwegian Beamline at ESRF with the wavelength $\lambda=0.6199$ \AA. By averaging over the symmetric equivalent reflections, 131 reflections are obtained for the refinement with FULLPROF$^{\rm S}$ \cite{rodriguez_1993s}. Fig. \ref{fig:refinement} presents the comparison between the observed and calculated intensities for the MnSc$_2$S$_4$ structure with Fd{\=3}m space group. The refined $x$ position for the S site is 0.2574(1) and the agreement factors are $R_{f}$ = 1.92 and $R_{f2}$ = 3.16. No Mn-Sc anti-site disorder is detectable, confirming the good quality of our single crystals. 

\textbf{Spherical neutron polarimetry matrices}

Spherical neutron polarimetry (SNP) matrices were measured with two configurations: one with the (HK0) plane as the horizontal scattering plane; another with the (HHL) plane as the horizontal scattering plane. SNP allows access to the complete nine elements of the polarization matrix $P_{ij}$ ($i$, $j$ = $x$, $y$, $z$), where $\hat{x}$ is opposite to the momentum transfer direction, $\hat{z}$ perpendicular to the horizontal scattering plane, and $\hat{y}$ forms a right-hand coordinate system. Tab.~\ref{tab:SNP_exp} lists SNP matrices measured with MuPAD on TASP at $T=1.36$ and 1.83 K for the two commensurate phases. The intensity of the [$\overline{0.75}$ 3.25 0] reflection at 1.83 K is too weak for SNP measurements. 

Tab.~\ref{tab:SNP_calc} lists SNP matrices calculated for the helical, cycloidal, and sinusoidally modulated collinear structures. The neutron polarization ratio of 96\% for benders before and after the sample is included in the calculation. A half-half ratio is assumed for the two helical (cycloidal) domains with opposite chiralities. The helical and collinear structures reproduce the 1.36 and 1.83 K data, respectively, while the cycloidal structure fits neither of them.

\begingroup
\squeezetable
 \begin{table*} 
 \caption{\label{tab:SNP_exp} SNP matrices measured with MuPAD on TASP.}
 \begin{ruledtabular}
 \begin{tabular}{c c c}
 Reflections & $T=1.36$ K & $T=1.83$ K \\
 \hline
 \multicolumn{3}{c}{(HK0) as the horizontal scattering plane}\\
 \hline
 \vspace{1mm}
$(0.75\ 0.75\ 0)$ & 
 $\left( \begin{array} {ccc} 
	-0.85(3) & -0.08(4) & 0.01(4) \\
    -0.07(4) & 0.12(4) &-0.01(4) \\    
     0.07(4) & 0.03(4) &-0.04(4)   
 \end{array} \right)$ &
 $\left( \begin{array} {ccc} 
	-0.90(3) & -0.04(4) & -0.19(4) \\
    -0.01(4) & 0.92(3) & 0.04(4) \\    
     0.02(4) & 0.02(4) &-0.86(3)  
 \end{array} \right)$ \\
 $(\overline{0.75}\ 3.25\ 0)$ &
  $\left( \begin{array} {ccc} 
	-0.86(5) & -0.13(4) & 0.21(5) \\
     0.12(4) & -0.38(4) & -0.02(4) \\    
     0.22(4) & -0.03(4) & 0.33(3)   
 \end{array} \right)$ &
 Intensity is too weak for SNP measurements\\
 \hline
 \multicolumn{3}{c}{(HHL) as the horizontal scattering plane}\\
 \hline
 \vspace{1mm}
  $(0.75\ 0.75\ 0)$ & 
 $\left( \begin{array} {ccc} 
	-0.90(2) & -0.05(3) & 0.13(3) \\
     0.13(2) & -0.04(2) & 0.01(2) \\    
     0.22(2) &  0.01(2) & 0.09(2)   
 \end{array} \right)$ &
 $\left( \begin{array} {ccc} 
	-1.00(5) & -0.09(6) & 0.16(5) \\
     0.12(6) & -0.89(5) & 0.01(5) \\    
     0.09(5) &  0.05(5) & 0.89(4)   
 \end{array} \right)$ \\
 $(\overline{0.25}\ \overline{0.25}\ 1)$ &
  $\left( \begin{array} {ccc} 
	-0.98(7) & -0.01(7) & 0.11(7) \\
    -0.01(7) & -0.70(8) &-0.09(7) \\    
    -0.02(6) & -0.02(7) & 0.54(8)   
 \end{array} \right)$ &
   $\left( \begin{array} {ccc} 
	-0.93(6) &  0.06(6) & 0.22(6) \\
     0.09(6) & -0.95(6) & 0.10(6) \\    
     0.18(5) &  0.19(6) & 0.85(6)   
 \end{array} \right)$
 \end{tabular}
 \end{ruledtabular}
 \end{table*}

\begingroup
\squeezetable
 \begin{table*}
 \caption{\label{tab:SNP_calc} SNP matrices calculated for the helical, cycloidal, and the collinear structures.}
 \begin{ruledtabular}
 \begin{tabular}{c c c c}
 Reflections & Helical & Cycloidal & Collinear\\
 \hline
 \multicolumn{4}{c}{(HK0) as the horizontal scattering plane}\\
 \hline
 \vspace{1mm}
$(0.75\ 0.75\ 0)$ & 
 $\left( \begin{array} {ccc} 
	-0.884 & 0 & 0  \\
     0 & 0.000  & 0 \\    
     0 & 0  & 0.000   
 \end{array} \right)$ &
 $\left( \begin{array} {ccc} 
	-0.884 & 0 & 0  \\
     0 & 0.884  & 0 \\    
     0 & 0  & -0.884     
 \end{array} \right)$ &
 $\left( \begin{array} {ccc} 	
	-0.884 & 0 & 0  \\
     0 & 0.884  & 0 \\    
     0 & 0  & -0.884  
 \end{array} \right)$\\
 $(\overline{0.75}\ 3.25\ 0)$ &
 $\left( \begin{array} {ccc} 
	-0.884 & 0 & 0  \\
     0 & -0.496 & 0 \\    
     0 & 0  & 0.496
 \end{array} \right)$ &
 $\left( \begin{array} {ccc} 
	-0.884 & 0 & 0  \\
     0 & 0.884  & 0 \\    
     0 & 0  & -0.884     
 \end{array} \right)$ &  
  $\left( \begin{array} {ccc} 
	-0.884 & 0 & 0  \\
     0 & 0.884  & 0 \\    
     0 & 0  & -0.884  
 \end{array} \right)$ \\
 \hline
 \multicolumn{4}{c}{(HHL) as the horizontal scattering plane}\\
 \hline
 \vspace{1mm}
  $(0.75\ 0.75\ 0)$ & 
 $\left( \begin{array} {ccc} 
	-0.884 & 0 & 0  \\
     0 & 0.000  & 0 \\    
     0 & 0  & 0.000  
 \end{array} \right)$ &
 $\left( \begin{array} {ccc} 
	-0.884 & 0 & 0  \\
     0 & 0.884  & 0 \\    
     0 & 0  & -0.884  
 \end{array} \right)$ &
  $\left( \begin{array} {ccc} 
	-0.884 & 0 & 0  \\
     0 & -0.884  & 0 \\    
     0 & 0  & 0.884  
 \end{array} \right)$ \\
 $(\overline{0.25}\ \overline{0.25}\ 1)$ &
 $\left( \begin{array} {ccc} 
	-0.884 & 0 & 0  \\
     0 & -0.707 & 0 \\    
     0 & 0  & 0.707  
 \end{array} \right)$ &
 $\left( \begin{array} {ccc} 
	-0.884 & 0 & 0  \\
     0 & -0.052  & 0 \\    
     0 & 0  & 0.052  
 \end{array} \right)$ &
  $\left( \begin{array} {ccc} 
	-0.884 & 0 & 0  \\
     0 & -0.884  & 0 \\    
     0 & 0  &  0.884  
 \end{array} \right)$ \\
  \end{tabular}
 \end{ruledtabular}
 \end{table*}

\textbf{Magnetization and susceptibility data}

The magnetization data were measured using a commercial superconducting quantum interference device (SQUID) magnetometer (Quantum Design MPMS-XL) equipped with a $^{3}$He cryostat in the Laboratory for Scientific Developments and Novel Materials, Paul Scherrer Institut, Villigen, Switzerland. The magnetic field was applied along the [001] direction.

Fig. \ref{fig:chi}a presents the temperature dependence of the susceptibility $\chi$ measured under different magnetic fields. With a small field of 0.1 T, $\chi$ shows a sudden drop at $T\sim1.7$ K, which overlaps with the incommensurate transition region revealed by the neutron diffraction experiments presented in the main text. This magnetization drop moves to lower $T$ with increasing field, suggesting a finite boundary for the helical phase.

In our magnetization measurements, a small jump is observed with increasing magnetic field. Fig. \ref{fig:chi}b shows the subtracted magnetization data $M_{diff} = M_{meas}-M_{fit}$, where $M_{fit}$ is obtained from a linear fit for the region $H$ = 4 $\sim$ 6 T. The position of the magnetization jump decreases with increasing $T$, which is compatible with the downwards shift observed in the $\chi(T)$ measurements.

\begin{figure} [h]
\includegraphics[width=0.48\textwidth]{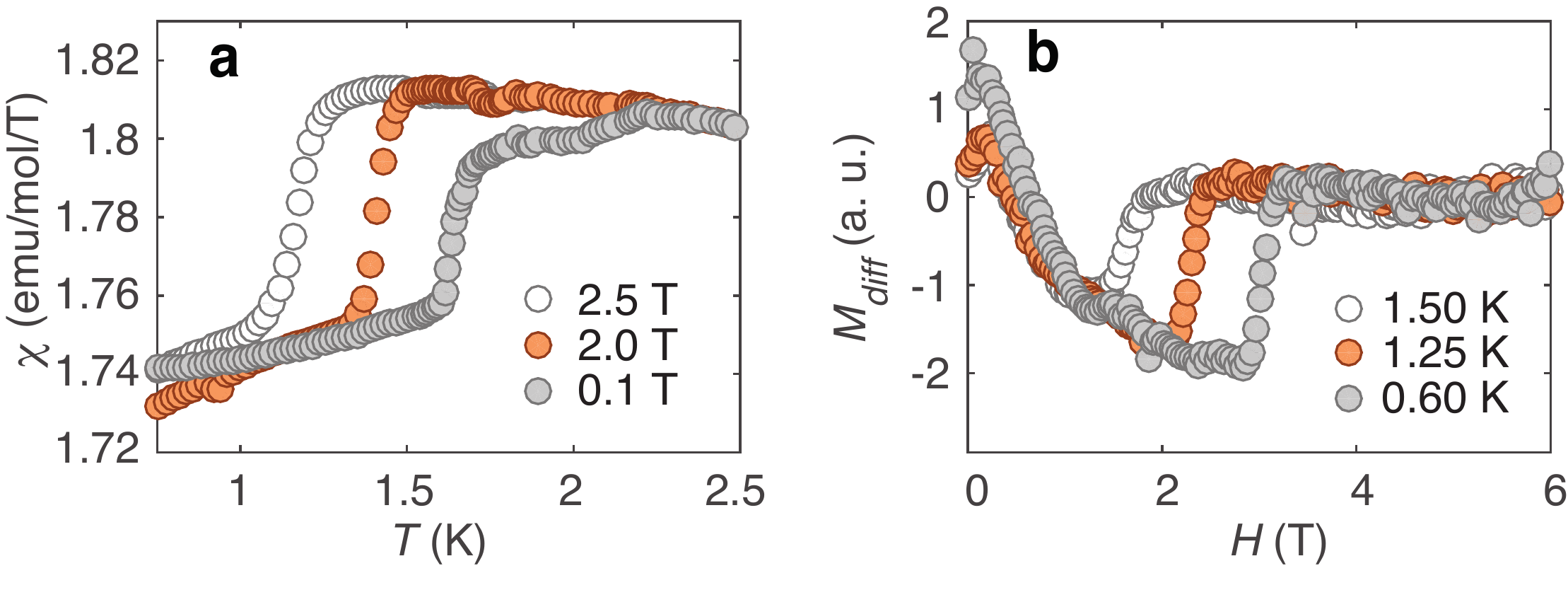}
\caption{\textbf{Magnetization and susceptibility data with field along the [001] direction. (a)} Temperature dependence of the magnetic susceptibility $\mathit{\chi}$ measured during warming. \textbf{(b)} Field dependence of the subtracted magnetization $M_{diff}$ (see text). \label{fig:chi}} 
\end{figure}

\begin{figure}
\includegraphics[width=0.46\textwidth]{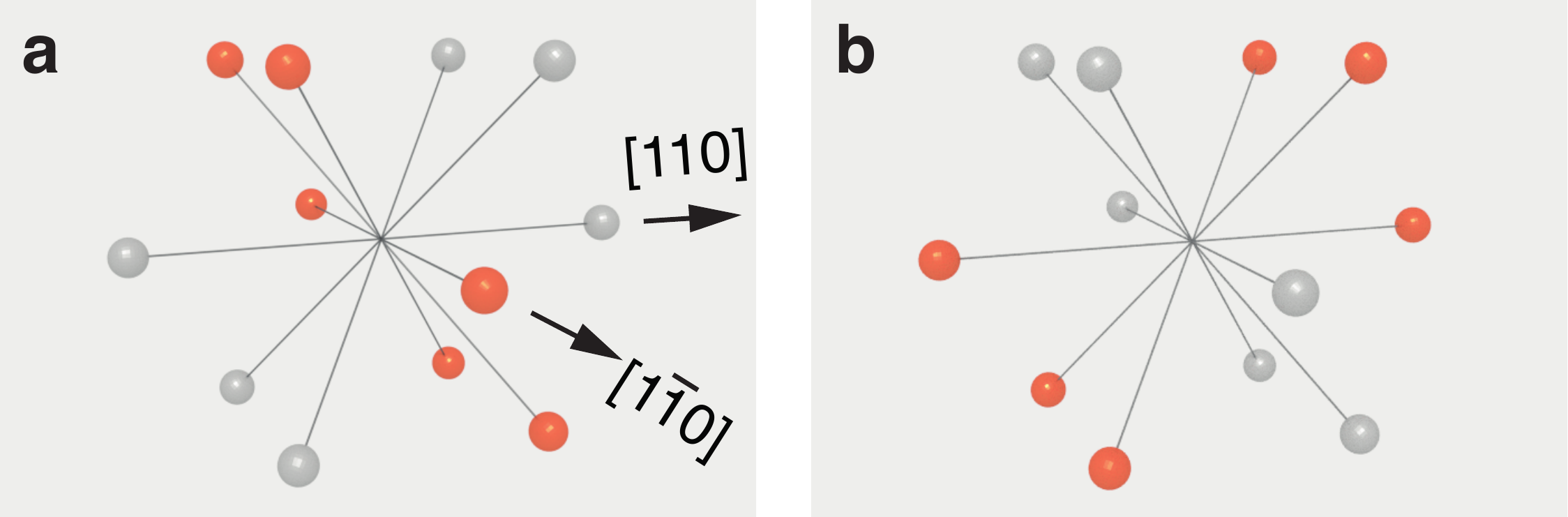}
\caption{\textbf{Comparison of the domain effects of the triple-\textbf{q} and collinear structures under a [111] magnetic field. (a)} Intensity distribution observed in the triple-\textbf{q} phase, which is a replica of Fig. 3e of the main text. \textbf{(b)} Intensity distribution expected for single-\textbf{q} collinear structures.\label{fig:compare}} 
\end{figure}

\textbf{Origin of the incommensurate phase}

Alternative explanations for the incommensurability observed in the transition region of 1.46 K $<T<1.64$ K may exist. One possibility might be that due to the competition of the exchange and increasing dipolar interactions, domains with collinear and helical structures coexist and alternate. Their adjustment might results in stripes with long incommensurate periodicity$^{\rm S}$ \citep{vindigni_2008, pregelj_2015s}. 

The existence of further-neighbour couplings might also play a role. Using mean-field calculations, we find that the addition of a relatively weak anti-ferromagnetic $J_4$ with $|J_4/J_3| \sim 1$ is able to reproduce the incommensurability observed along the [1\=10] direction. This incommensurate propagation vector might be locked to a nearby commensurate \textbf{q} position at lower temperature$^{\rm S}$ \cite{lee_2008s}, giving rise to the re-entrance transition into the commenusrate helical phase observed at 1.46 K.\\

\begin{figure} [h!]
\includegraphics[height=0.8\textwidth]{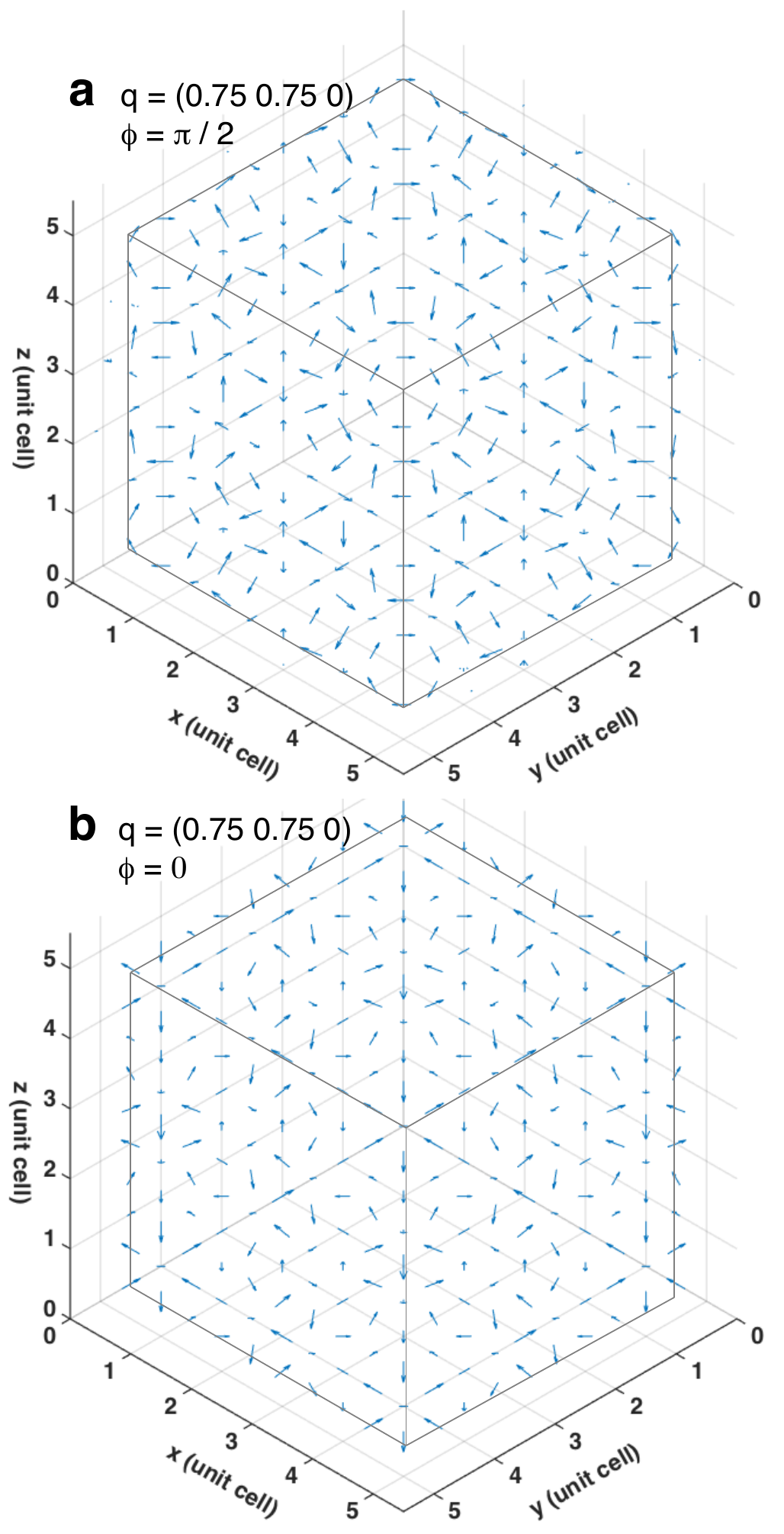}
\caption{\textbf{Triple-q structures preserving the $C_3$ symmetry viewed along the [111] direction. (a)} The triple-\textbf{q} structure for $\mathbf{q}=(0.75\ 0.75\ 0)$ with $\phi_1 = \phi_2 = \phi_3 = \pi/2$. \textbf{(b)} The triple-\textbf{q} structure for $\mathbf{q}=(0.75\ 0.75\ 0)$ with $\phi_1 = \phi_2 = \phi_3 = 0$. \label{fig:triple1}} 
\end{figure}

\begin{figure} [h!]
\includegraphics[height=0.8\textwidth]{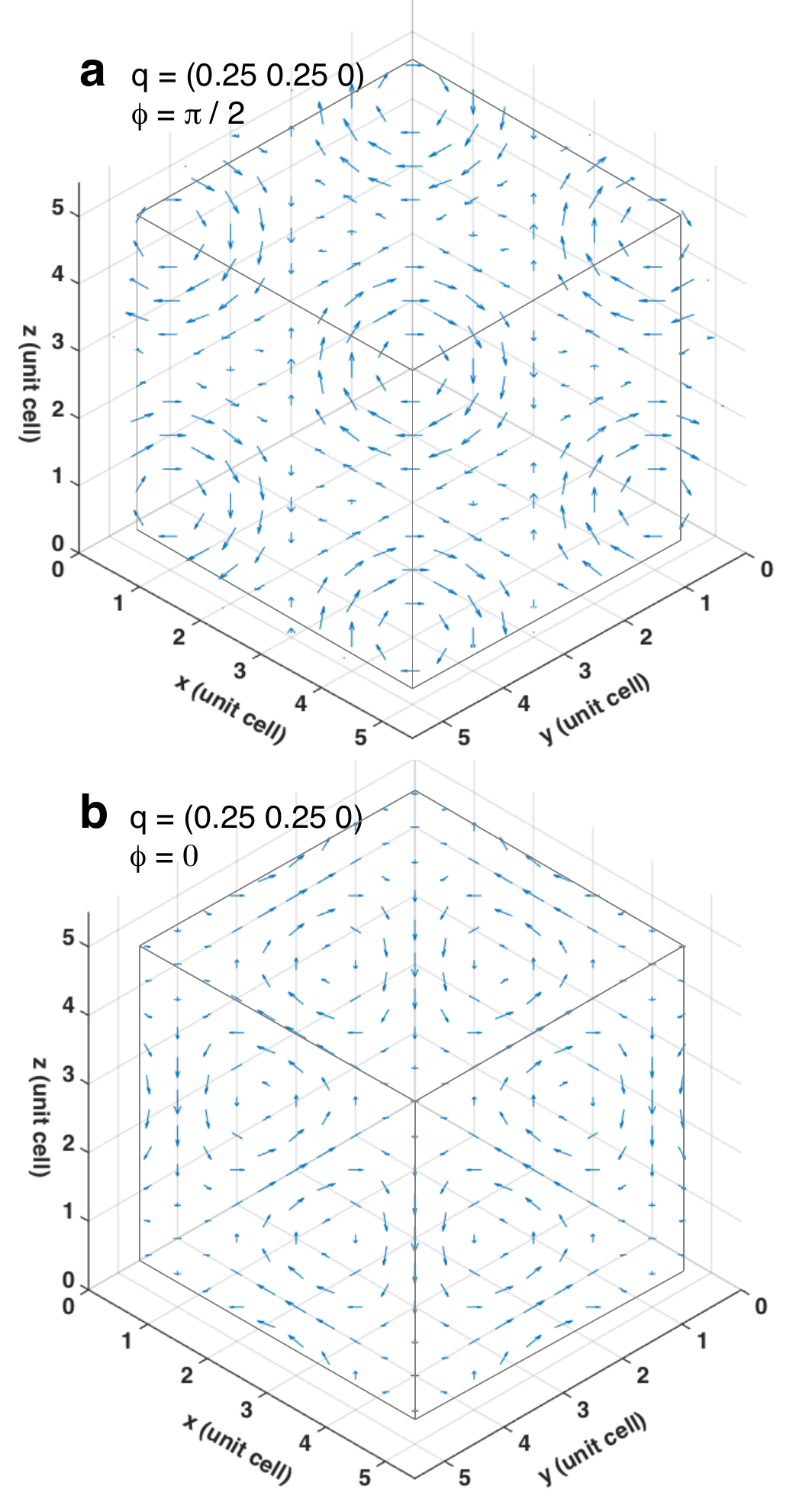}
\caption{\textbf{Triple-q structures expected for \textbf{q} = (0.25 0.25 0). (a)} \textbf{(a)} Expected triple-\textbf{q} structure for $\mathbf{q}=(0.25\ 0.25\ 0)$ with $\phi_1 = \phi_2 = \phi_3 = \pi/2$. \textbf{(b)} Expected triple-\textbf{q} structure for $\mathbf{q}=(0.25\ 0.25\ 0)$ with $\phi_1 = \phi_2 = \phi_3 = 0$. \label{fig:triple2}} 
\end{figure}

\textbf{Detailed analysis of domain effects}

\textbf{Collinear phase.}
Collinear structures have the highest magnetic susceptibility $\chi_\perp$ along the direction perpendicular to the spins. Therefore, a magnetic field will favor domains with spins perpendicular to the field direction. Under a magnetic field along the [001] direction, the favored domains in the MnSc$_2$S$_4$ collinear phase are those with \textbf{q} arms in the (HK0) plane. Our neutron diffraction measurement, which is summarized in Fig. 3b of the main text, confirms this domain re-distribution. The field-induced increase of the (0.75 0.75 0) intensity in the collinear phase shown in Fig. 4b of the main text can also be explained by this domain effect.

\textbf{Helical phase.}
Similar to collinear structures, helical structures also possess the highest magnetic susceptibility $\chi_\perp$ in the direction perpendicular to the helical plane. As a result, domains with the helical plane parallel with the magnetic field have a lower susceptibility $\chi_\parallel$ and are disfavored by the magnetic field. For the helical phase of MnSc$_2$S$_4$, these are domains with \textbf{q} arm inside of the (HK0) domain if the magnetic field is applied along the [001] direction. As is summarized in Fig. 3c of the main text, such a domain redistribution is confirmed experimentally. The vanishing of the (0.75 0.75 0) intensity shown in Fig. 4a of the main text further corroborates this domain effect. 

\textbf{Triple-\textbf{q} phase.}
Domain effects for the triple-\textbf{q} phase under both [001] and [111] magnetic fields have been illustrated in the main text. Here we provide further support for the triple-\textbf{q} structure by comparing its domain effect under a [111] field (shown in Fig. 3e of the main text and reproduced here as Fig. \ref{fig:compare}a) with that of the collinear structure. As in the case of a [001] field, if arms of the $\langle$0.75 0.75 0$\rangle$ star are treated independently, datasets collected under a [111] field ($T$ = 1.60 K and $H$ = 3.5 T, as that in Fig. 3e of the main text) can be refined very well with the sinusoidally modulated collinear structure. However, if arms were truly independent, the domain distribution expected for the collinear structure, which is shown in Fig.~\ref{fig:compare}b, will be completely opposite to our observations shown in Fig.~\ref{fig:compare}a. Thus the 12 arms of the $\langle$0.75 0.75 0$\rangle$ star are not independent, but couple together, constituting a triple-\textbf{q} structure as detalied in the main text.  \\

\textbf{Vortex lattices of the triple-\textbf{q} phase} 

\begin{figure}
\includegraphics[width=0.38\textwidth]{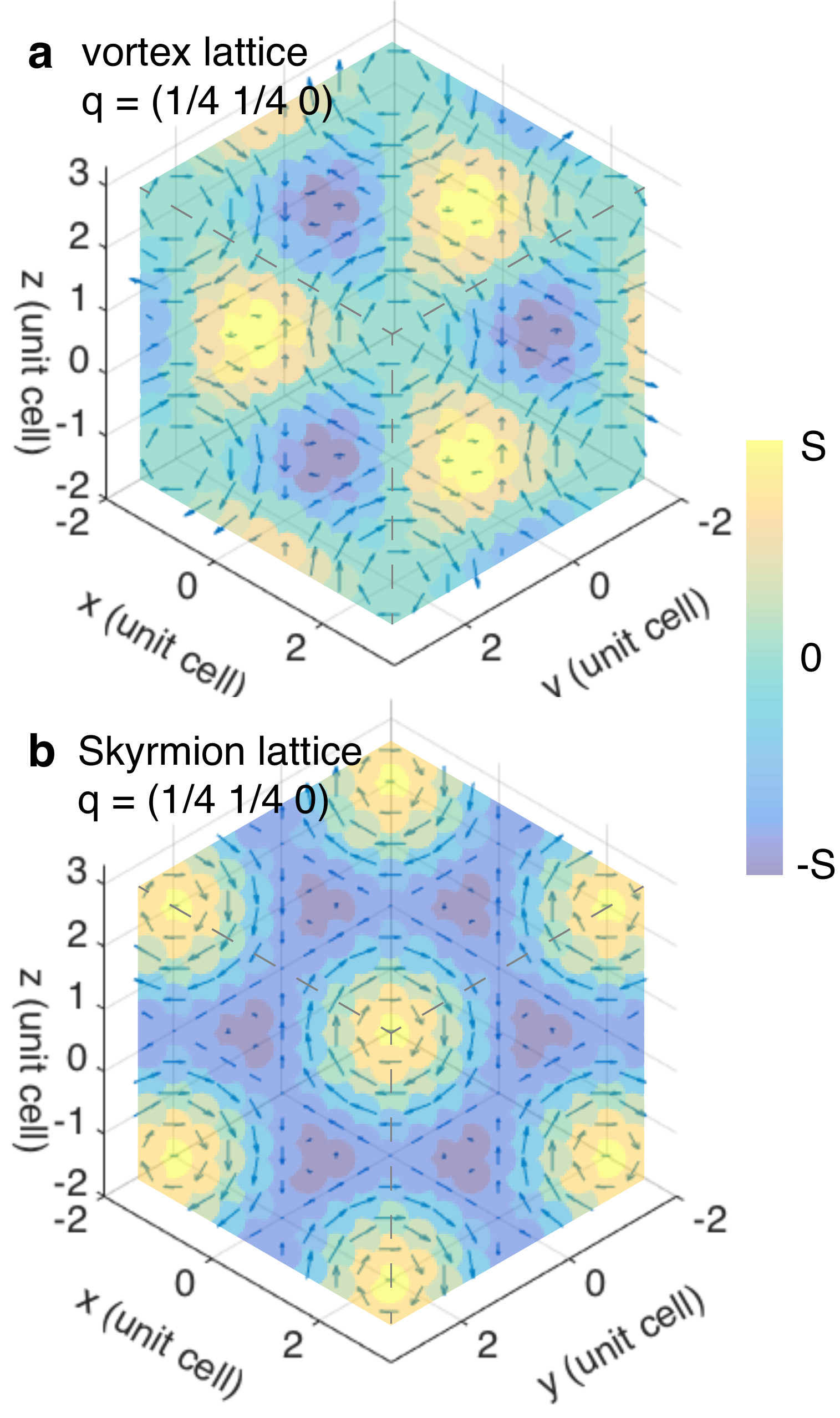}
\caption{\textbf{Comparison between vortex and skyrmion lattices for q = (0.25 0.25 0). (a)} The vortex lattice composed of one-dimensional basis vectors with $\phi_1 = \phi_2 = \phi_3 = \pi/2$, similar to that shown in Fig. 4c. \textbf{(b)} The skyrmion lattice composed of two-dimensional basis vectors perpendicular to the wave vector \textbf{q}. The colormap shows the size of the spin component $S_{\parallel}$ along the (111) direction. \label{fig:skyrmion}} 
\end{figure}

Fig.~\ref{fig:triple1}a,b present two possible triple-\textbf{q} structures that preserve the $C_3$ symmetry along the [111] direction. The phase factors are $\phi_1 = \phi_2 = \phi_3 = \pi/2$ in Fig. \ref{fig:triple1}a and $\phi_1 = \phi_2 = \phi_3 = 0$ in Fig. \ref{fig:triple1}b. Sites along the [111] direction possess the same moment size and direction. A winding feature around the $C_3$ axis is observed for the spin components perpendicular to the [111] direction. Due to the long propagation vector $\mathbf{q}=(0.75\ 0.75\ 0)$ in MnSc$_2$S$_4$, the periodicity in real space is very short, leading to strong variance between neighbouring spins. 

As a comparison, Fig.~\ref{fig:triple2}a,b show the corresponding triple-\textbf{q} structures expected for a shorter propagation vector $\mathbf{q}=(0.25\ 0.25\ 0)$, which might be realized in the \textit{A}-site spinels with a frustration ratio $|J_2/J_1|\sim0.15$. The evolution between neighbouring spins is more continuous. In the case of $\phi_1 = \phi_2 = \phi_3 = \pi/2$, vortices have a uniform chirality. While in the case of $\phi_1 = \phi_2 = \phi_3 = 0$, vortices with opposite chirality can be observed. \\

\textbf{Comparison with the skyrmion lattice}

Fig. \ref{fig:skyrmion} presents a comparison between vortex and skyrmion lattices expected for \textbf{q} = (0.25 0.25 0)$^{\rm S}$ \cite{nagaosa_2013s}. Although the spin components in the (111) plane $S_{\parallel}$ have the similar winding character in both structures, the out-of-plane components $S_{\perp}$ behave quite differently. In the skyrmion lattice, the basis vectors are two-dimensional, meaning that each arm represents a helical component. This helicity causes the $S_{\perp}$ components to increase when approaching the winding axis$^{\rm S}$ \cite{nagaosa_2013s}. In contrast, in the vortex lattice, due to its one-dimensional basis vectors, $S_{\perp}$ components become zero when approaching the winding center. \\

\textbf{Transition region between the collinear and triple-q phases}

The transition region between the collinear phase and triple-\textbf{q} phases needs further investigation. As is shown in Fig. 4c of the main text, this transition region exists in the same temperature window as the incommensurate phase observed under zero field, suggesting a possible connection between them. Presently we find that the $T$-dependence of the intensity ratio between the (0.75 0 0.75) and the (0.75 0.75 0) arms is non-monotonous, which is summarized in Fig. \ref{fig:nonmono}. The neutron diffraction datasets collected for both arms can be refined with the same collinear structure, similar to the situation in the triple-\textbf{q} phase.  

\begin{figure} [h!]
\includegraphics[width=0.18\textwidth]{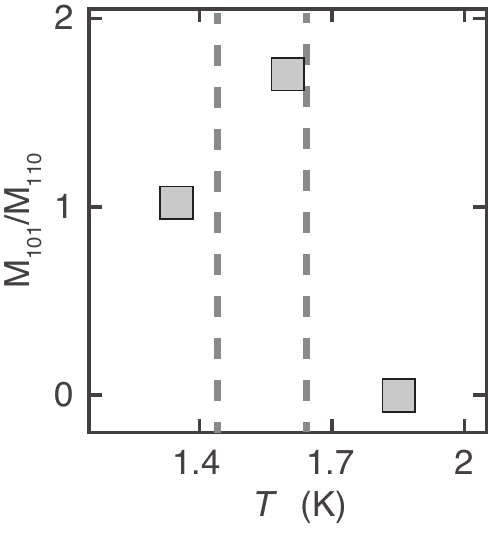}
\caption{\textbf{Non-monotonous evolution from the collinear phase to the triple-\textbf{q} phase. (a)} Weight ratio between (0.75 0 0.75) and (0.75 0.75 0) arms: $M_{101}/M_{110} \doteq \sqrt{I_{101}/I_{110}}$, where $I$ is the observed integrated intensity and M is the refined moment size. Measurements were performed under a magnetic field of 3.5 T at $T=1.82$, 1.65, and 1.38 K. Dashed lines indicate the phase boundaries of the collinear and triple-\textbf{q} phases.\label{fig:nonmono}} 
\end{figure}


%

\end{document}